\documentclass{iopart}

\usepackage{color}
\usepackage[usenames,dvipsnames,svgnames,table]{xcolor}
\usepackage{ulem}

\usepackage{iopams}  
\usepackage{mathrsfs}
\usepackage{graphics}
\renewcommand{\bs}{\boldsymbol}
\renewcommand{\mr}{\mathrm}
\renewcommand{\ms}{\mathscr}
\newcommand{\mc}{\mathcal}
\newcommand{\f}[2]{\frac{#1}{#2}}
\newcommand{\df}[2]{\frac{\partial #1}{\partial #2}}
\newcommand{\kron}[2]{ \delta^{#1}_{\phantom{#1}#2}}
\newcommand{\ie}{{\it i.e. }}
\newcommand{\eg}{{\it e.g. }}
\newcommand{\chri}[2]{\Gamma^{#1}_{#2}}
\newcommand{\p}{\phantom}
\newcommand{\eqref}[1]{(\ref{#1})}
\newcommand{\noi}{\noindent}
\renewcommand{\etal}{{\it et al.~}}

\begin{document}

\title[The Bondi--Sachs metric at the vertex of a null cone]
{The Bondi--Sachs metric at the vertex of a null cone: 
axially symmetric vacuum solutions }

\author{Thomas M\"adler$^{1,\,2}$ and Ewald M\"uller$^1$ }
\address{$ ö1$ Max Planck Institute for Astrophysics, 
Karl Schwarzschild Str. 1, D-85741 Garching, Germany, \\
$^2$ Laboratoire Univers et Theorie (LUTH), 
Observatoire de Paris, CNRS, Universit\'e Paris Diderot, 
5 place Jules Janssen, 92190 Meudon, France} 

\ead{thomas.maedler@obspm.fr} 

\begin{abstract} 
In the Bondi--Sachs formulation of General Relativity space-time is
foliated via a family of null cones.  If these null cones are defined
such that their vertices are traced by a regular world-line then the
metric tensor has to obey regularity conditions at the vertices.
We explore these regularity conditions when the world line is a time-like
geodesic. In particular, we solve the Einstein equations for the
Bondi--Sachs metric near the vertices for axially symmetric vacuum
space-times.  The metric is calculated up to third order corrections
with respect to a flat metric along the time-like geodesic, as this is
the lowest order where non-linear coupling of the metric coefficients
occurs. We also determine the boundary conditions of the metric to
arbitrary order of these corrections when a linearized and axially
symmetric vacuum space-time is assumed. In both cases we find that (i)
the initial data on the null cone must have a very rigid angular
structure for the vertex to be a regular point, and (ii) the initial
data are determined by functions depending only on the time of a
geodesic observer tracing the vertex.  The latter functions can
  be prescribed freely, but if the vertex is assumed to be regular
  they must be finite and have finite derivatives along the geodesic.
\end{abstract} 
\maketitle
\section{Introduction}
The pioneering work of Bondi  and Sachs led to a rigorous understanding
of gravitational waves at large distances from a compact source (Bondi \etal
1962, Sachs 1962). They used a coordinate chart
$(x^0,\,x^1,\,x^2,\,x^3) =(u,\,r,\theta,\,\phi)$ for the metric that
was adapted to out-going null cones and where only one coordinate, the
areal distance $r$, varies along the out-going null geodesics
generating these cones. The other coordinates are the `retarded time'
$u$, and the standard spherical angles $\theta$ and $\phi$. Assuming a
Minkowskian observer at infinitely far distances from an isolated
gravitating object they showed that such a system can lose mass
(energy) only via gravitational radiation. Their results were
confirmed by Newman  and Unti (1962) who used the Newman--Penrose
formalism (Newman  and Penrose 1962, 1963) to demonstrate the advantage
of null foliations in General Relativity in order to characterize gravitational
waves.

An observer at null infinity can only keep track of what is happening
in the future or past of a source that emits gravitational
waves. However, if one is interested in studying the emission process
itself, e.g. by means of numerical simulations, one needs to consider
conditions {\it at} the source.  Tamburino  and Winicour (1966) were the
first to impose the boundary conditions for the gravitational field at
a finite distance on the null cone, \ie they placed a Minkowskian
observer on a time-like world-tube of finite space-like radius, \eg
the radius of a star.  This world-tube null formalism, where out-going
null geodesics are attached to a world-tube, forms the basis of many
numerical codes that are currently used to solve the Einstein
equations in a domain extending from a finite-size world-tube to null
infinity (see Winicour 2012a, for a recent review).

When the radius of the world-tube becomes zero, the world-tube
degenerates into a world-line, \ie into the time-like curve of an
inertial observer tracing both the vertices of out-going null cones
and the origin of the coordinate system. This so-called Fermi observer
is a non-rotating observer in a rectangular Minkowskian coordinate
system along the world-line. Isaacson \etal (1981) integrated the
Einstein equations for the Bondi metric (Bondi 1960) when the
world-tube had zero radius. The boundary conditions for the Bondi
metric were chosen such that it approaches flat space values at the
origin. To assure this behavior they restricted the fall-off of the
metric functions towards $r=0$ to certain positive powers of $r$. The
work of Isaacson \etal motivated G{\'o}mez \etal (1994) and Siebel
\etal (2002) to solve numerically the vacuum Einstein and the
Einstein-fluid equations for the Bondi metric in axial symmetry from
the vertex to null infinity. In their integration schemes, the boundary
conditions for the Bondi metric included the lowest order curvature
contribution. Both used a correct, but {\it ad hoc}, ansatz for the
boundary conditions near the vertex, which they did not provide for
all variables, however.

In this work we systematically investigate the boundary conditions  of
the Bondi--Sachs metric near the vertex.
The vertex of a null cone is the focal point where all null geodesics
generating the cone either converge to or emanate from. From the mathematical point of view, the null cone is not differentiable at its vertex, and consequently derivatives of the metric and therefore the curvature tensor cannot be calculate there. In particular, any tensor  tensor that is expressed in terms
of coordinates, like the Bondi--Sachs coordinates, adapted to the null cone cannot be expanded in terms of
a Taylor series with respect to these coordinates at the vertex. Therefore, the boundary conditions for any fields expressed in Bondi--Sachs coordinates are {\it  a priori} not known at the vertex. In principle these boundary conditions can be chosen freely. However, the vertex is the origin of an inertial observer along the geodesic, who implies regular boundary conditions. The assumption of such an observer at the vertex also requires that  the vertex is a {\it regular point} for the considered tensor fields on the null cone (Dautcourt, 1967). A regular point $\ms{P}$ (the vertex) for a function $f$ (\eg a metric component) on a open domain $\ms{U}$ (\ie the null cone without the vertex) is defined as a point on the boundary $\partial \ms{U}$,  where $f$ has a Taylor expansion based on $\ms{P}$ into $\ms{U}$ (Freitag and Busam, 2005). If the vertex is a regular point in the null cone, we will call it a {\it regular vertex}. 

To find the proper boundary conditions of fields on a null cone at its vertex, we have to make additional assumptions:

\begin{enumerate}
  \item Assuming that the vertices of the null cones are traced by a world-line of a Fermi observer, we require this world line to be a time-like geodesic since then the acceleration of the   curve vanishes, which greatly simplifies the analysis. This geodesic should be a {\it regular time-like geodesic}, \ie its tangent vector should nowhere vanish along the curve. 
  \item The geodesic should be a contained in a {\it convex normal neighborhood} (Hawking and Ellis, 1973). Such a neighborhood, $\Gamma$ say, has the property that two distinct points in $\Gamma$ can be connected by a unique geodesic and  at any point, $\ms{P}$, in $\Gamma$ Riemann normal coordinates, $z^\mr{a}$, can be established such that the metric in the neighborhood of $\ms{P}$ can be written as
\begin{equation}
\label{ }
g_{\mr{ab}} = \eta_{\mr{ab}}- \f{1}{3}\mc{R}_{\mr{acbd}}({\ms{P}})z^\mr{c}z^\mr{d} +...\;\;,
\end{equation}
where $\eta_{\alpha\beta} = \mr{diag}(-1,\,+1,\,+1,\,+1)$ and $\mc{R}_{\mr{acbd}}({\ms{P}})$ is the Riemann tensor evaluated at $\ms{P}$. In this regard,
a metric is called  a {\it regular metric} at a point $\ms{P}\in\Gamma$, if it is invertible at $\ms{P}$ and its determinant is finite and nonzero at $\ms{P}$. 
\end{enumerate}

\noi
These two assumption are crucial to find the boundary conditions of the Bondi--Sachs metric at the vertex of the null cone. These two assumptions are necessary to find the boundary conditions.  Assuming a regular metric of an inertial observer moving along a time-like geodesic, we will use  coordinate transformations to find a Bondi--Sachs metric at the time-like geodesic. This Bondi--Sachs metric will not be regular at the vertex because the angular base vectors are not defined there.  Nevertheless, by this procedure we will obtain the desired boundary conditions for the Bondi--Sachs metric corresponding to an observer in an inertial frame at the vertex, because the initial metric was a regular metric.  
To this end we will address four questions:
\begin{enumerate}
  \item[(I)] What are the generic requirements of the boundary conditions of the Bondi--Sachs
    metric near the vertex of the null cone if the vertex is a regular point in the null cone?
  \item[(II)] How can these  boundary conditions be determined?
  \item[(III)] What are the explicit  boundary conditions  for the
    Bondi--Sachs metric of an axially symmetric vacuum space-time?
  \item[(IV)] What do the regular boundary conditions imply for the initial
    data on a null cone with a regular vertex for axially symmetric vacuum space-times?
\end{enumerate}
To find the answers to these questions, we construct a Fermi normal coordinate system (Misner and Manasse, 1963) along the time-like geodesic. Based on this rectangular coordinate system, we then calculate a metric on a null cone where the radial coordinate is an affine parameter. This null metric is subsequently transformed to a Bondi--Sachs metric, by changing the radial coordinate over to an areal distance coordinate.   This has the advantage that the Einstein equation split into a  hierarchy of hypersurface equations and evolution equations (Bondi \etal, 1962, Winicour, 2012a) which simplifies their numerical treatment on a null cone. If  an affine parameter is used instead, one of the hypersurface equations contains an additional time-derivative which is numerically more challenging (Winicour, 2012b).
 Alternatively, we could  start with a regular metric along the time-like geodesic  in deDonder coordinates as in Thorne (1980), Thorne and Hartle  (1985),  Zhang (1986) and  calculating the boundary conditions for a  Bondi--Sachs metric via an appropriate coordinate transformation. Since these  authors derived their metric expansions for  vacuum space-times, only,  we do not uses their expansions to answer question (I), since assuming a vacuum space-time restricts the generic properties of the boundary conditions (see Sec. \ref{sec:direct_fermi}). Second (and  more importantly) starting from a Fermi normal system is simpler because  the  Fermi normal coordinates are geometrically adapted to the properties of the convex normal neighborhood. We will later compare our results with those of Zhang (1986), who obtains an expansion of the metric in the $3+1$ formulation of space-time.

The behavior of arbitrary fields near the vertices of null cones was
first discussed by Penrose (1963), who realized that these
fields should be expanded in terms of normal coordinates defined at
the point of origin of the expansion.  His argument was formal
providing only a sketch of the expansion. Friedrich (1986)
considered the expansion for the case of the conformal vacuum
Einstein equations, and, like Penrose, used the spin frame formalism
to derive the connection and the curvature variables.  As we are
interested in the conditions at the vertex of the null cone in the
Bondi--Sachs formulation,  where the curvature is given as a
function of the metric,  the method of Penrose (1963) and
  Friedrich (1986) does not directly apply.  

Ellis \etal (1985) discussed regularity conditions at the vertices of
past null cones which are traced by a time-like geodesic of a
cosmological (comoving) observer.  Their work motivated Poisson  and
 collaborators (Poisson, 2004, 2005, Poisson \etal 2006a, 2006b, 2010, 2011) to use light cone coordinates at a world-line to
obtain the metric of a non-rotating black hole moving along the world
line.  As both Ellis \etal and Poisson  and Vlasov used an affine
parameter instead of an areal distance along along the null rays, they
would have had to transform the former into the latter by an
appropriate coordinate transformation to obtain the Bondi--Sachs
metric.

In principle, we could have used the
results of Ellis \etal and Poisson \etal as a starting point to
answer questions (I)-(IV), but our aim is to present a complete and pedagogical derivation of the boundary conditions of a Bondi--Sachs metric at the vertex from a regular metric of an inertial observer along a time-like geodesic. With this in mind, we follow the approach of Ellis \etal (1985, App. A), but we provide additional pedagogical steps in the derivation and generalization of their results. Regarding Poisson and collaborators, we do not need the machinery of a bi-tensor theory or the Synge world function (Synge, 1960)  to find a null metric with an affine parameter as radial coordinate.  However, we will recover the results of Ellis \etal and Poisson \etal in our calculations.

Choquet--Bruhat \etal (2010, 2011) and Chru{\'s}iel  and Jezerski (2010)
investigated the Cauchy problem of General Relativity for initial data
given on the null cone, and Chru{\'s}ciel  and Paetz (2012) took these
studies up comparing the results with those of Friedrich (1986).  They
analyzed the `vertex problem' raising questions about uniqueness and
existence of solutions. Although these results are mathematically very
interesting, they are too general to help in formulating boundary 
conditions for a Bondi-Sachs metric, and in answering question (III) and (IV) above.  We note that regularity issues at the
origin of a curvi-linear coordinate system are also of importance in
the 3+1 formulation of General Relativity. The first in-depth analysis
of this problem for axially symmetric systems is due to Bardeen  and
Piran (1983).

The article is organized as follows: After defining our notation
  in the reminder of the introduction, we derive in Section
  \ref{sec:BStrafo} the limiting behavior of a Bondi--Sachs metric
from a regular Fermi normal coordinate system, and answer
  question (I) concerning the generic properties of the Bondi--Sachs
  metric. In Section \ref{sec:sols}, we discuss and respond to question
  (II)-(III). In Section \ref{sec:SumDiss}, we conclude with a summary and discussion of the results, and
  we respond to question (IV).
\subsection{Notation}
We assume the existence of a smooth, four-dimensional manifold
$\ms{M}^4$ with a Lorentzian metric $g$ and signature
$(-,\,+,\,+,\,+)$. The flat metric is always denoted with
$\eta$. Greek indices ($\alpha,\,\beta,\,...$) and the small Latin
indices ($\mr{a,\,b,...,f}$) run from $0$ to $3$, while the small
Latin indices ($i,j,k, ...$) can assume the values $1,\;2$, or $3$,
respectively. Capital Latin indices $(A,\,B,\,C\,...) $ label
angular directions on a sphere and take the values 2 or 3 for the
coordinates $(\theta, \phi)$ or $(y= -\cos\theta,\, \phi)$,
respectively. The Kronecker symbol is
$\delta_{ij} = \mr{diag}(+,\,+,\,+)$. Unless
stated otherwise, the Einstein summation convention holds. We use
five distinct symbols for the four-dimensional space-time
coordinates: (i) $y^\alpha$ are  arbitrary 
coordinates;
(ii) $z^\mr{a}$ are Riemann normal coordinates;
(iii) $y^\mr{a}$ are Fermi normal coordinates;
(iv) $\bar{x}^\alpha$ are affine null coordinates; and (v)
$x^\alpha$ are Bondi--Sachs coordinates. Partial derivatives are
denoted by a comma, and covariant derivatives by the $\nabla-$symbol.
For the Christoffel symbols $\chri{\alpha}{\mu\nu}$, the Riemann
$\mc{R}^\alpha_{\p{\alpha}\beta\mu\nu}$ and Ricci tensor
$\mc{R}_{\mu\nu}$, we follow the convention of Misner \etal
(1972). Moreover, we use geometrized units where the speed of light
and the gravitational constant are equal to unity. Associated Legendre
polynomials of the first kind, $P^m_l(y)$, are defined as in Jackson
(1999), \ie 
\begin{displaymath} 
    \fl P^m_l(y) = \f{(-1)^m}{2^l l!}(1-y^2)^{m/2} 
                   \f{d^{l+m}}{dy^{l+m}}(y^2-1)^l\;\;.
\end{displaymath}
\section{A Bondi--Sachs metric derived from a regular metric}
\label{sec:BStrafo}
\subsection{A regular metric in Fermi normal coordinates}
Let $(\ms{M}^4,\,g)$ be the four-dimensional space-time. Suppose
$(\ms{M}^4,\,g)$ contains a simply convex normal neighborhood
$\Gamma$ with a regular time-like geodesic $\bs{c}(\tau)$, where $c^\alpha$ is
its unit, time-like tangent vector and $\tau$ the proper time. Since $\Gamma$ is a convex normal neighborhood and $\bs{c}(\tau)$ is contained in $\Gamma$, a regular metric $g_{\alpha\beta}$ can be found along $\bs{c}(\tau)$. Let $y^\alpha$ be a set of  four-dimensional arbitrary coordinates along $\bs{c}(\tau)$ such that the metric $g_{\alpha\beta}$ is regular and  can be expanded along $\bs{c}(\tau)$ like 
\begin{equation}
\label{eq:general_exp}
\/\fl g_{\alpha\beta}(y^\mu) =
g_{\alpha\beta}\Big|_{\bs{c}(\tau)}(y^\mu) + 
g_{\alpha\beta,\kappa_1}\Big|_{\bs{c}(\tau)} y^{\kappa_1} + 
\f{1}{2}g_{\alpha\beta,\kappa_1\kappa_2}\Big|_{\bs{c}(\tau)}\;\;
y^{\kappa_1}y^{\kappa_2} +...\;\;,
\end{equation}
where the  coefficients are evaluated along $\bs{c}(\tau)$.
Because of arguments given in section \ref{sec:BSlemma} below, it is
sufficient to consider the expansion only up to the quadratic term.

At any point on $\bs{c}(\tau)$ we choose an orthonormal tetrad
$\bs{e_{\mr{a}}}(\tau)$ which is parallel propagated along
$\bs{c}(\tau)$, \ie $\nabla_{\bs{c}(\tau)}\bs{e_{\mr{a}}}=0$.  The
time-like base vector of the coordinate system $\bs{e_{\mr{0}}}$ is
tangent to $\bs{c}(\tau)$, \ie $\bs{e_0}(\tau)
=\partial/\partial\tau$.  On any point of the geodesic $\bs{c}(\tau)$,
we send out space-like geodesics $b(\tau,\,\bs{n}, \ell)$ which are
parametrized by an affine parameter $\ell$ and point into the
direction $\bs{n} = n^{\mr{i}} \bs{e_{\mr{i}}}(\tau)$, \ie $n^0 =0$
and $\bs{n} =\partial/\partial \ell |_{\bs{c}(\tau)}$.  The affine
parameter $\ell$ of the geodesics $b(\tau,\,\bs{n}, \ell)$ is defined
to be zero along $\bs{c}(\tau)$.  This specifies the coordinates
$y^\alpha$ as Fermi normal coordinates $y^\mr{a}$ (Misner  and Manasse, 1963, Ni  and Li,
1979),  which are given by
\begin{equation}
\label{eq:FNC}
 \fl y^0:=        \tau\;\;,\qquad
     y^{\mr{i}}: = \ell n^{\mr{i}}\;\;, \qquad
\ell(y^\mr{i}):=       \sqrt{ \delta_{\mr{ij}}y^{\mr{i}}y^{\mr{i}}}\;\;,
\end{equation}
where $dn^{\mr{i}}/d\tau=0$ and $n^{\mr{i}} = y^{\mr{i}}/\ell$ are the
direction cosines of the space-like geodesics $b(\tau,\,\bs{n},
\ell)$.  Note that the three-dimensional coordinates $y^i$ are Riemann
normal coordinates
\footnote{For a thorough discussion of Riemann normal coordinates, see
  \eg Schouten (1954), Thomas (1991), and Iliev (2006).}
for every value of $\tau$ on $\bs{c}(\tau)$.  Constructing the
coordinates in that way assures that the metric along $\bs{c}(\tau)$
is the Minkowski metric, because the base vectors are orthonormal
along $\bs{c}(\tau)$. In addition, it implies that the first order
partial derivatives, $ g_{\alpha\beta,\kappa_1}$, vanish along
$\bs{c}(\tau)$, because of the parallel transport equation of
$n^\mr{i}$ along $\bs{c}(\tau)$ and the geodesic equations of
$\bs{c}(\tau)$ and $b(\tau,\,\bs{n}, \ell)$.  The particular form of
the second partial derivatives at the geodesic $\bs{c}(\tau)$ was
first derived by Misner  and Manasse (1963), and the corresponding
metric reads
\numparts
\begin{eqnarray}
\fl g_{00}(y^{\mr{a}}) & = & -1 - \mc{R}_{\mr{0i0j}}\Big|_{\bs{c}(\tau)}(\tau)y^{\mr{i}}y^{\mr{j}} + \mc{O}([y^{\mr{a}}]^3)\;\;,\\
\fl g_{\mr{0k}} (y^{\mr{a}}) & = & -\f{2}{3} \mc{R}_{\mr{0ikj}}\Big|_{\bs{c}(\tau)}(\tau)y^{\mr{i}}y^{\mr{j}} + \mc{O}([y^{\mr{a}}]^3)\;\;,\\
\fl g_{\mr{km}} (y^{\mr{a}}) & = & \delta_{\mr{ij}}-\f{1}{3} \mc{R}_{\mr{ikjm}}\Big|_{\bs{c}(\tau)}(\tau)y^{\mr{i}}y^{\mr{j}}+ \mc{O}([y^{\mr{a}}]^3)\;\;,
\end{eqnarray}
\endnumparts 
where $\mc{R}_{\mr{abcd}}$ are the Riemann normal components of a
Fermi observer along the time-like geodesic $\bs{c}(\tau)$.  These
components are space-time invariants, \ie knowing all twenty
independent Riemann normal components along $\bs{c}(\tau)$ allows one
to construct uniquely (up to spatial rotations) a metric up to quadratic terms of a power series expansion with respect to Fermi normal coordinates along the time-like geodesic in a
sufficiently small neighborhood $\Gamma$ of the  geodesic.
\subsection{An affine null metric}
\label{sec:affine}
The null coordinate $\bar{x}^0:=\bar{\tau}_w$ labels null cones whose
vertices are along the time-like geodesic $\bs{c}(\tau)$, where
$\bar{\tau}_w$ is equal to the proper time $\tau$ along the geodesic,
\ie $\bar{\tau}_w|_{\bs{c}(\tau)}= \tau$. For points not on
$\bs{c}(\tau)$, $\bar{\tau}_w$ is constant along parametrized null
geodesics that emanate from $\bs{c}(\tau)$.  We introduce the three
spatial coordinates $(\bar{x}^1,\, \bar{x}^2,\, \bar{x}^3)$. The
radial coordinate $\bar{x}^1:=\ell$ is a positive affine parameter for
null rays emanating from $\bs{c}(\tau)$, where $\ell=0$. The two
coordinates $\bar{x}^A:=(\bar{x}^2,\,\bar{x}^3)$, which are constant
both along the time-like geodesic $\bs{c}(\tau)$ and the null
geodesics, label the direction angles of the null rays.  In the
following, we will use the notation $\ms{O}$ when we refer to an
arbitrary point on $\bs{c}(\tau)$ that is also the vertex of an
arbitrary null cone.
For $\ell$ to be an affine parameter, it has to obey the condition
\begin{equation}\label{eq:def_affine_p}
\fl  k^\mr{a} \nabla_\mr{a} \ell = 1\;\;,
\end{equation}
where $\nabla_\mr{a}$ is the covariant derivative with respect to the metric in Fermi normal coordinates. Equation \eref{eq:def_affine_p}
holds, if we define $\ell$ as in \eqref{eq:FNC} and choose the
null vector ${k^\mr{a}(y^\mr{b}):=(1,\,y^\mr{i}/\ell)}$ in the Fermi normal
coordinate system.  Then $\ell$ does not only measure the space-like
distance from the geodesic $\bs{c}(\tau)$ in a Fermi frame, but it
also gives the positive affine distance between a vertex $\ms{O}$ and
points on the respective null cone of $\ms{O}$. We note that an affine
parameter along a curve is generally defined only up to the
transformation $\lambda=a\ell+b$, where $a$ and $b$ are arbitrary
constants. However, our choice $\ell=0$ on $\bs{c}(\tau)$ implies
$b=0$. From $a=\mr{sign}(a)|a|$ and the definition of $\ell$
\eqref{eq:FNC} one sees that multiplying the Fermi normal coordinates
$y^\mr{i}$ by $|a|$ corresponds to scaling these coordinates, \ie
without loss of generality we can set $|a|=1$. Hence, null rays
$\bs{k}(\lambda)$ emanating from $\bs{c}(\tau)$ are most generally
parametrized with the affine parameter $\lambda = w\ell$, where
$w=\pm1$. We choose $w=1$ for future-pointing null rays along $\bs{c}(\tau)$ and $w=-1$ for
past-pointing ones by imposing the normalization condition
\begin{equation}\label{eq:norm_k}
\fl \lim_{\ell \rightarrow 0}  c_\mr{a} k^\mr{a} = - w\;\
\end{equation}
along $\bs{c}(\tau)$, where $c^\alpha$ is the tangent vector of
$\bs{c}(\tau)$.
 
A null geodesic $\bs{k}(\lambda)$ in the null-cone $\bar{\tau}_w =
const$ emanating in $\bar{x}^A-$direction, admits an expansion in
Fermi normal coordinates $y^\mr{a}$ with respect to the affine
parameter $\lambda$ of the form (using $\lambda=w\ell$ and $w^2=1$ )
\begin{eqnarray}\label{eq:null_geodesic_taylor}
\fl{y}^{\mr{a}}(\bar{x}^\mu) 
& =& 
\bar{\tau}_w \kron{\mr{a}}{{\tau}}  
+w\ell k^{\mr{a}}\Big[y^\mr{b}(\bar{x}^A)\Big]
+\f{\ell^2}{2!}F^\mr{a}\Big[y^\mr{b}(\bar{\tau}_w,\,\bar{x}^A)\Big]
 +\f{w\ell^3}{3!}G^\mr{a}\Big[y^\mr{b}(\bar{\tau}_w,\,\bar{x}^A)\Big]
 +\mc{O}(\ell^4)\nonumber,\\
 \fl
\end{eqnarray}
where the coefficient functions are evaluated along $\bs{c}(\tau)$. Setting $\ell=0$ in \eqref{eq:null_geodesic_taylor} shows that it
describes the time-like geodesic $\bs{c}(\tau)$ with the tangent
vector $c^\alpha=\kron{\alpha}{\tau}$, and that the tangent vector of
the null geodesics $k^\mr{a}=(1/w)d y^\mr{a} / d\ell|_{\bs{c}(\tau)}$
does not depend on $\bar{\tau}_w$, because the $\bar{x}^A$ are
constant along $\bs{c}(\tau)$. If $\bs{c}(\tau)$ were no geodesic, an
additional dependence of $\bar{\tau}_w$ through the directional vector
$k^\mr{a}$ must be taken into account (Newman  and Posadas, 1969).
Hereafter, we use the null vector $k^\mr{a}$ in the form $k^\mr{a} =
\big(1,\,n^\mr{i}(\bar{x}^A)\big)$, where $n^\mr{i} $ is a
three-dimensional unit vector being parametrized by the angles
$\bar{x}^A$.

\noindent
The geodesic equations of the null rays $\bs{k}(w, \ell)$ read in Fermi
normal coordinates
\begin{equation}\label{eq:geodesic_k}
\fl \f{d^2y^\mr{a}}{d\ell^2}=-\chri{\alpha}{\mr{bc}}(y^\mr{d})\f{dy^{\mr{b}}}{d\ell}\f{dy^{\mr{c}}}{d\ell}\;\;,
\end{equation}
and further differentiation with respect to $\ell$ gives
\begin{eqnarray}\label{eq:geodesic_k_diff}
\fl \f{d^3y^{\mr{a}}}{d\ell^3}&=&-\chri{\mr{a}}{\mr{bc,d}}(y^\mr{g})\f{dy^\mr{b}}{d\ell}\f{dy^\mr{c}}{d\ell}\f{dy^{\mr{d}}}{d\ell}\;\;.
\end{eqnarray}
Inserting \eqref{eq:null_geodesic_taylor} into \eqref{eq:geodesic_k}
and \eqref{eq:geodesic_k_diff}, and equating the thus obtained
relations along $\bs{c}(\tau)$ by setting $\ell=0$ leads to
\begin{eqnarray*}
\fl F^\mr{a}
 (\bar{\tau}_w,\bar{x}^A) & = & 0 \;\;,\\
\fl G^\mr{a}
 (\bar{\tau}_w,\bar{x}^A) & = & -\chri{\mr{a}}{\mr{bc,d}}\Big|_{\bs{c}(\tau)}(\bar{\tau}_w)k^b(\bar{x}^A)k^c(\bar{x}^A)k^d(\bar{x}^A)\;\;.
\end{eqnarray*}
In Fermi normal coordinates $G^\mr{a}$ vanishes, because the
derivative $\chri{\mr{a}}{\mr{bc,d}}$ can be expressed as a linear
function of the Riemann tensor $\mc{R}^{\mr{a}}_{\mr{bcd}}$, which is
antisymmetric in the last two indices, and these two indices are
contracted symmetrically.  Hence, the parametric representation of the
null rays $\bs{k}(w, \ell)$ that emanate from the time-like geodesic
$\bs{c}(\tau)$ reads in the Fermi frame
\begin{displaymath}
\fl y^a(\bar{x}^\alpha) = \bar{\tau}_w\kron{\mr{a}}{\bar{\tau}_w}  +w\ell k^{\mr{b}}(\bar{x}^A) +\mc{O}(\ell^4)\;\;.
\end{displaymath}
The metric in affine null coordinates,  $\bar{x}^\alpha$, is given by the
coordinate transformation
\begin{displaymath}
\fl g_{\alpha\beta}(\bar{x}^\mu) = 
    g_{\mr{ab}}\Big[y^\mr{c}(\bar{x}^\mu)\Big]\df{y^\mr{a}}{\bar{x}^\alpha}
    (\bar{x}^\mu)\,\df{y^\mr{b}}{\bar{x}^\beta}(\bar{x}^\mu)\;\;,
\end{displaymath}
and the Jacobian $(\partial y^\mr{a}/\partial\bar{x}^\alpha)$ between $y^\mr{a}$
and $\bar{x}^\alpha$ by
\numparts
\begin{eqnarray}
\fl \df{{y}^\mr{a}}{\bar{\tau}_w} (\bar{x}^\mu)& = & \kron{\mr{a}}{{\tau}} 
  +\f{\ell^2}{2!}F^\mr{a}_{,\bar{\tau}_w}(\bar{\tau}_w,\,\bar{x}^A)
 +\f{w\ell^3}{3!}G^\mr{a}_{,\bar{\tau}_w}(\bar{\tau}_w,\,\bar{x}^A)
  +\mc{O}(\ell^4)\;\;,
 \\
\fl \df{{y}^\mr{a}}{\ell}(\bar{x}^\mu) & = &w k^\mr{a}(\bar{x}^A)
      +\mc{O}(\ell^4)\;\;,
 \\ 
\fl \df{{y}^\mr{a}}{\bar{x}^A}(\bar{x}^\mu) & = & w\ell k^\mr{a}_{,A}(\bar{x}^B)
  +\f{\ell^2}{2!}F^\mr{a}_{,A}(\bar{\tau}_w,\,\bar{x}^B)
 +\f{w\ell^3}{3!}G^\mr{a}_{,A}(\bar{\tau}_w,\,\bar{x}^B)
   +\mc{O}(\ell^4)\;\;,
\end{eqnarray}
\endnumparts
which contains the derivatives of the coefficients $F^\mr{a}$ and
$G^\mr{a}$ with respect to the null coordinates.  Although these
coefficients vanish at $\ell=0$, their derivatives are not necessarily
zero there, \ie we have to calculate the derivatives first, and
subsequently evaluate them in the limit $\ell=0$.  Using the first
derivatives of the Christoffel symbols in a Fermi normal coordinate
system as in Ni  and Li (1979), we calculated the metric
$g_{\alpha\beta}(\bar{x}^\alpha)$ in affine null coordinates near the
time-like geodesic $\bs{c}(\tau)$:
\numparts
\begin{eqnarray}
\fl g_{00}(\bar{\tau}_w, \,\ell,\,\bar{x}^A) & = & -1 -\ell^2\mc{A}(\bar{\tau}_w,\,\bar{x}^A)
+\mc{O}(\ell^3)\label{eq:AN_g00}\;\;, \\
\fl g_{0A}(\bar{\tau}_w, \,\ell,\,\bar{x}^A)&=&-\f{2}{3}w\ell^3\mc{B}_{A}(\bar{\tau}_w,\,\bar{x}^A)
+\mc{O}(\ell^4)\;\;,\\
\fl g_{1\mu}(\bar{\tau}_w, \,\ell,\,\bar{x}^A) & = & -w\kron{0}{\mu}\;\;,\\
\fl g_{AB}(\bar{\tau}_w, \,\ell,\,\bar{x}^A)&=& \ell^2q_{AB}(\bar{x}^C) -\f{\ell^4}{3}\mc{S}_{AB}(\bar{\tau}_w,\,\bar{x}^A)
       +\mc{O}(\ell^5)\label{eq:AN_gAB}\;\;,
\end{eqnarray}
\endnumparts 
where $q_{AB}(\bar{x}^C) = \delta_{\mr{ij}}n^\mr{i}_{,A}n^\mr{j}_{,B}$
is a unit sphere metric around $\ms{O}$, because $n^\mr{i}(x^A)$ is a
unit vector labeling points on a sphere, and  the coefficients 
\numparts
\begin{eqnarray}
\fl \mc{A}(\bar{\tau}_w,\,\bar{x}^C) &=& \mc{R}_{\mr{0a0b}}\Big|_{\bs{c}(\tau)}(\bar{\tau}_w)k^\mr{a}(\bar{x}^C)k^\mr{b}(\bar{x}^C)\label{eq:contractions_riem_A}\;\;,\\
\fl \mc{B}_A(\bar{\tau}_w,\,\bar{x}^C) &=& \mc{R}_{\mr{0abc}}\Big|_{\bs{c}(\tau)}(\bar{\tau}_w)k^\mr{a}(\bar{x}^C)k^\mr{b}_{,A}(\bar{x}^C)k^\mr{c}(\bar{x}^C)\label{eq:contractions_riem_B}\;\;,\\
\fl \mc{S}_{AB}(\bar{\tau}_w,\,\bar{x}^C) &=& \mc{R}_{\mr{abcd}}\Big|_{\bs{c}(\tau)}(\bar{\tau}_w)k^\mr{a}(\bar{x}^C)k^\mr{b}_{,(A}(\bar{x}^C)k^\mr{c}(\bar{x}^C)k^\mr{d}_{,B)}(\bar{x}^C)\label{eq:contractions_riem_S}
\end{eqnarray}
\endnumparts
depend on the Riemann normal components $\mc{R}_{\mr{abcd}}
\big|_{\bs{c}(\tau)} (\bar{\tau}_w)$ evaluated along the
geodesic $\bs{c}(\tau)$, \ie $\mc{R}_{\mr{abcd}} \big|_{\bs{c}(\tau)}
(\bar{\tau}_w) = \mc{R}_{\mr{abcd}}(\tau_w)$. We point out that the
components $g_{1 \mu}$ are exact, and as they are equal to a
constant, they do not require an order symbol. In particular,
$|g_{10}|=1$, because $\ell$ is an affine parameter.

The components of the affine null metric \eqref{eq:AN_g00}-
\eqref{eq:AN_gAB} obtained by us agree with those in the literature
(Ellis \etal 1985, Poisson \etal 2010).  While Ellis \etal (1985)
calculated a null metric at a past null cone only, we derived the
metric for points both in a past and future null cone, which is
achieved by setting $w=1$ (future) or $w=-1$ (past) in our solution.
The difference between our work and that of Poisson \etal (2010)
concerns the computational approach of how to find the affine null
metric along $\bs{c}(\tau)$. Poisson \etal (2010) use the Synge world
function (Synge, 1960) to obtain the metric, whereas we use an
explicit coordinate transformation based on a regular Fermi normal
coordinate system. Both approaches are legitimate and it is a matter
of taste which one prefers.
\subsection{The Bondi--Sachs metric}
\label{sec:BSlemma}
Using the affine null metric $g_{\mu\nu}(\bar{x}^\alpha)$ derived in
the previous subsection, we calculate a Bondi--Sachs metric and the
corresponding Bondi--Sachs coordinates are labeled by $x^\alpha$.
The null coordinate $x^0=\tau_w$ and the angular coordinates $x^A$ are
defined like their barred counterparts, $\bar{\tau}_w$ and
$\bar{x}^A$, respectively. The radial coordinate $x^1:=r$ is a
positive areal distance coordinate, \ie surfaces $d\tau_w=const$ and
$dr=const$ have the area $4\pi r^2$. This also requires that the
determinant of the 2-metric $g_{AB}$ divided by $r^4$ does not vary
with time and radius.

The areal distance $r$ is defined by (Jordan \etal (1961), Sachs,
1961, Newman and Penrose,1963)
\begin{equation}
\label{eq:def_area_dist}
\fl k^\mu\nabla_\mu  r = r \Theta(k) \;\;,
\end{equation} 
where $\Theta(k):=\f{1}{2}\nabla_\mu k^\mu$ is the expansion rate of
null rays with tangent vector $k^\mu$. Inserting the null vector
$k^\mu(\bar{x}^\alpha) = w\kron{\mu}{\ell}$ into equation
\eqref{eq:def_area_dist} we obtain
\begin{equation}
\label{eq:sol_r_BS}
\fl r(\bar{x}^\mu) =\ell\Big(1-\f{\ell^2}{3}\mc{S}(\bar{\tau}_w,\,\bar{x}^A)+\mc{O}(\ell^3) \Big)^{1/4} \;\;, \quad \mc{S}(\bar{\tau}_w,\,\bar{x}^A):=q^{AB}(\bar{x}^A)S_{AB}(\bar{\tau}_w,\,\bar{x}^A)\,.
\end{equation}
The equation shows that the affine parameter is equal to the areal
distance $r$ both for $|\ell| \approx 0$ and in a flat space-time. The
latter holds, because for a flat metric $\mc{S}$ vanishes due to its
dependence on the Riemann tensor.

We define a coordinate transformation between the affine null
coordinates $\bar{x}^\mu$ and the Bondi--Sachs coordinates $x^\mu$
according to
\begin{equation}
\label{ }
\fl
\tau_w = \bar{\tau}_w\;\;,\qquad
r(\bar{x}^\mu) = \ell -\f{\ell^3}{12}\mc{S}(\bar{\tau}_w,\,\bar{x}^A)
                 +\mc{O}(\ell^4)\;\;,\qquad
x^A=\bar{x}^A\;\;
\end{equation}
Applying this coordinate transformation to the metric components
\eqref{eq:AN_g00} - \eqref{eq:AN_gAB} results in the following
Bondi--Sachs metric near the geodesic $\bs{c}(\tau)$:
\numparts
\begin{eqnarray}
\fl g_{00}(x^\mu)   &=& -1 - r^2\mc{A}(\tau_w,\,x^C) 
                       + \mc{O}(r^3) \;\;,
\label{eqBS_from_AN_uu}\\
\fl g_{0A}(x^\mu)   &=& -\f{2}{3}wr^3\mc{B}_A(\tau_w,\,x^C)
                       + \mc{O}(r^4)\;\;,\\
\fl g_{1\mu} (x^\mu) &=& -w\Big[1+ \f{r^2}{4}\mc{S}(\tau_w,\,x^C) 
                       + \mc{O}(r^3)\Big]\kron{0}{\mu}\;\;, \\
\fl g_{AB}(x^\mu)   &=& r^2q_{AB}(x^C) -
                       \f{r^4}{6}\Big[\mc{S}_{AB}(\tau_w,\,x^C) - 
                       \f{1}{2}q_{AB}(x^C)\mc{S}(\tau_w,\,x^C)\Big]  
                       + \mc{O}(r^5)\;\;.
\label{eqBS_from_AN_AB}
\end{eqnarray}
\endnumparts 
Note that the above metric is indeed of the Bondi--Sachs type, because
the determinant of $g_{AB}/r^2$ is a function of $x^A$, only.

Traditionally, the Bondi--Sachs metric is written in the form (Bondi
1960, Sachs, 1962)
\footnote{ (1) Bondi and Sachs, and also most numerical relativists
  (see Winicour (2012) for a review) use retarded time $u:=\tau_{+1}$
  as time coordinate; (2) for computational convenience, we renamed
  the original metric function $V/r$ as $\exp(2\Phi+2\beta)$ in the
  metric component $g_{00}(x^\alpha)$. }
\begin{equation}
\label{eq:BS_trad}
\fl ds^2  = -\rme^{2\Phi+4\beta}d\tau_w^2-2w\rme^{2\beta}d\tau_w dr + 
            r^2 h_{AB}\Big(dx^A-U^Ad\tau_w\Big) 
                     \Big(dx^B-U^Bd\tau_w\Big)\;\;,
\end{equation}
where the determinant of the 2-metric, $\det(h_{AB})$, is a function
of $x^A$ only.  Because of this restriction, the tensor $h_{AB}$ possesses only two degrees of freedom. Using
the standard spherical coordinates as angular coordinates,
$x^A=(\theta,\,\phi)$, allows us to write $h_{AB}$ in the form (van
der Burg, 1966)
\begin{equation}
\label{eq:vdB}
\fl h_{AB}dx^Adx^B =\rme^{ 2\gamma}\cosh(2\delta)\,d\theta^2 + 
                    2\sin\theta\sinh(2\delta)d\theta d\phi + 
                   \rme^{ -2\gamma} \cosh(2\delta) \sin^2\theta \,d\phi^2\;\;.
\end{equation}
By linearizing  $h_{AB}$ with respect to $\gamma$ and $\delta$, it can be seen that $\gamma$ and $\delta$ correspond to the two components of a two-dimensional transverse and trace-less tensor, \eg the tensor $\chi_{AB}$ in Chu\'sciel \etal (1998, 2002). 
According to \eqref{eq:BS_trad} the Bondi--Sachs metric
$g_{\alpha\beta}(x^\mu)$ is not defined at $r=0$, because the
four-dimensional volume element $\sqrt{-g}$ vanishes there.

For convenience we introduce the quantity $\ms{F}\in\{\gamma,\,
\delta,\, \beta,\, \Phi,\, U^A\}$ to abbreviate the set of the six
functions describing the Bondi--Sachs metric. To find the behavior of
$\ms{F}$ at the vertex, we assume that $\ms{F}$ can formally be
expanded into a power series at $r=0$, \ie
\begin{equation}
\fl \ms{F}(x^\mu) = \ms{F}^{(0)}(\tau_w,\,x^A) + 
                   r\ms{F}^{(1)}(\tau_w,\,x^A) +
                   \f{r^2}{2!}\ms{F}^{(2)}(\tau_w,\,x^A)+ \ldots\,,
\end{equation}
\noi where the coefficient functions $\ms{F}^{(n)}(\tau_w,\,x^A)$ are evaluated along the
geodesic $\bs{c}(\tau)$. Inserting the expansion of each element of
$\ms{F}$ into (\ref{eq:BS_trad}) and (\ref{eq:vdB}) results in a
series expansion for every component of the Bondi--Sachs metric in
terms of the elements of $\ms{F}$. A comparison of these series
expansions with the expression given in equations
(\ref{eqBS_from_AN_uu}) - (\ref{eqBS_from_AN_AB}) yields the following
relations between $\ms{F}$ and the contracted Riemann normal
components $\mc{A},\,\mc{B}_A$ and $\mc{S}_{AB}$:
\numparts
\begin{eqnarray}
\fl  \gamma(x^\alpha)&=& -\f{r^2}{12}\f{ \mc{S}_{\theta\theta}(\tau_w,\,x^A) \sin^2\theta - \mc{S}_{\phi\phi}(\tau_w,\,x^A) }{\sin^2\theta}
+ \mc{O}(r^3)\;\;,\label{eq:BS_F_gam}\\
\fl  \delta(x^\alpha)&=&  -\f{r^2}{6}\f{\mc{S}_{\theta\phi}(\tau_w,\,x^A)}{\sin\theta}+\mc{O}(r^3)\;\;,\\
\fl  \beta(x^\alpha) & =& \;\; \;\;\;\f{r^2}{8}\mc{S}(\tau_w,\,x^A)+\mc{O}(r^3)\;\;,\\
\fl U^A(x^\alpha) & =& -\f{2}{3} r q^{AC}\mc{B}_C(\tau_w,\,x^D) +\mc{O}(r^2)\;\;, \\
\fl \Phi(x^\alpha)&=& \f{r^2}{2}\Big[\mc{A}(\tau_w,\,x^D)-\f{1}{2}\mc{S}(\tau_w,\,x^D)\Big]+\mc{O}(r^3)\;\;. \label{eq:BS_F_Phi}
\end{eqnarray}
\endnumparts
Equations (\ref{eq:BS_F_gam})-(\ref{eq:BS_F_Phi}) show how the
traditionally-used Bondi--Sachs metric functions $\ms{F}$ behave to
lowest order near the vertex of the null cone, when the vertex
coincides with the origin of a Fermi normal coordinate system along
$\bs{c}(\tau)$. Hereafter, these lowest order corrections are also
referred to as first-order corrections, $\ms{C}_1$, of the Bondi--Sachs metric with
respect to the flat metric at the vertex along the time-like geodesic or briefly first-order
corrections of the Bondi-Sachs metric. These first order corrections $\ms{C}_1$ arise at $\mc{O}(r)$ for $U^A$  and at $\mc{O}(r^2)$ for $\beta,\,\gamma,\,\delta$, and $\Phi$, where as higher-order correction coefficients $\ms{C}_n$ are expected at  $\mc{O}(r^n)$ for $U^A$  and at $\mc{O}(r^{n+1})$ for $\beta,\,\gamma,\,\delta$, and $\Phi$, respectively.  

Since equations (\ref{eq:BS_F_gam})-(\ref{eq:BS_F_Phi}) were derived form a regular metric at the vertex and the vertices were assumed to be on a timeline geodesic, we  deduce (without solving the Einstein equations) the following regularity requirements on a power series expansion of  the  traditionally-used functions $\ms{F}$, which assure that the Bondi--Sachs metric is regular at its origin:
\begin{enumerate}
  \item[$(i)$] The power series of the metric functions $\ms{F}$ in $r$ must
    start at $r=0$ with a certain positive power of $r$, \ie  $\gamma,\delta,\,\beta,\,\Phi$ are of $\mc{O}(r^2)$ and $U^A$ of $\mc{O}(r)$. 

  \item[$(ii)$] The angles $x^A$ must parameterize topological spheres centered at the vertex and the radial coefficients of $\ms{F}$ must
    show a specific angular behavior determined by the
    contractions of the Riemann normal components\footnote{Physically speaking, this requirement means that if the space-time $(\ms{M}^4, g)$ is curved, the path of a null ray emanating from the vertex is affected in the neighborhood of the vertex (\ie the Fermi observer) by the curvature (the Riemann tensor and its covariant derivatives) at the vertex. } with the tangent $c^\alpha$, the null vector $k^\alpha$, and $k^\alpha_{,A}$ along $\bs{c}(\tau)$. 
    
   \vspace{1ex}
    In Sec. \ref{sec:vac_sol}, we will show that the topological requirement is crucial for regularity, since it is also possible to have non-regular solutions at the vertex, when this assumption is     
    not imposed.      
    
  \item[$(iii)$] The radial expansion of $\ms{F}$  have specific numerical factors in the
    radial expansion coefficients, for example the factor $1/12$ in \eref{eq:BS_F_gam}. 
  
\end{enumerate}

\noi
Two additional important properties of the functions $\ms{F}$ at the
vertex cannot be inferred from the first-order corrections in
\eqref{eqBS_from_AN_uu}-\eqref{eqBS_from_AN_AB}, because they manifest
themselves only at second- or third-order deviations from the flat
metric at $r=0$.  While the first-order corrections originate from the
second derivative of the Fermi metric at the origin through a
coordinate transformation, the second and third-order corrections
result from the third and fourth derivative of the Fermi metric.  To
learn more about the two missing properties we consider a qualitative
argument, which also avoid tedious calculations of the corresponding
coordinate transformations. We recall that in Fermi normal coordinates
the Christoffel symbols vanish along $\bs{c}(\tau)$, and that in any
coordinate system, where the Christoffel symbols vanish, the following
correspondence can be made between third and fourth partial
derivatives of the metric and the Riemann tensor (Schouten, 1954):
\numparts
\begin{eqnarray}
\fl   \partial^3 g &=&L\Big[ \nabla (\mr{Riem}) \Big]\;\;,\\
\fl   \partial^4 g &=& L\Big[\nabla^2 (\mr{Riem}),\; (\mr{Riem})^2\Big]\;\;,\label{eq:Riem_quad_fermi}
\end{eqnarray} 
\endnumparts 
where $\partial^n g$ is the $n^{th}$--partial derivative of the
metric, $L(\cdot)$ a linear functional of its arguments, $(\mr{Riem})$
the Riemann normal components, and $\nabla^n$ the $n^{th}$--covariant
derivative, respectively. The explicit dependence between the
third and fourth-order partial derivatives of the metric and the
Riemann tensor in Fermi normal coordinates is given in Ni  and Li
(1979), Dolgov \etal (1983), and Ishii \etal (2005).

To calculate the Bondi--Sachs metric including the second and
third-order corrections, the corresponding affine null metric must be
determined first.  Following and extending schematically the procedure
of section \ref{sec:affine} to higher order in the approximation
reveals that the affine null metric involves time derivatives of the
Riemann normal components at the vertices on $\bs{c}(\tau)$.  These
derivatives arise because covariant derivatives of the Riemann tensor
in the Jacobian are contracted, like \eg $k^\mr{a}\nabla_\mr{a} \mc{A}
= k^{\rm{a}}k^{\mr{b}}k^{\mr{c}}\nabla_{\rm{a}}\mc{R}_{\mr{0b0c}}$
with the null vector $k^\alpha$, which is given by $k^\mr{a}(x^A) =
\kron{\mr{a}}{\tau_w} + n^i(x^A)\kron{\mr{a}}{\mr{i}}$.  Starting at
second-order corrections, these time derivatives show up in
hierarchical manner, \ie the second-order corrections contain
first-order time derivatives of first-order corrections to the flat
metric at the vertex, like \eg $\mc{A}_{,\tau_w}$.  Similarly, the
third-order corrections contain second-order time derivatives of
first-order corrections, like \eg $\mc{A}_{,\tau_w\tau_w}$.  Moreover,
there are first-order time derivatives of second-order corrections in
the third-order correction coefficients, like \eg the time derivative
of the second-order correction $n^\mr{i}(x^A)\nabla_\mr{i}\mc{A}$.

At this stage, it is instructive to compare where time derivatives of the Riemann normal components occur in the metric expansion using different coordinate conditions. In table \ref{tab:comp_deD_FNC_AN}, we list the occurrence of the contracted Riemann normal components $\mc{R}_{0i0j}$ and their spatial covariant derivatives in the first three correction coefficients of $g_{00}$ to a flat metric on the geodesic for an expansion  in Fermi-normal coordinates, in deDonder coordinates and in affine null coordinates, respectively. The behavior for the Fermi normal coordinates can be taken from Ni  and Li
(1979), Dolgov \etal (1983), or Ishii \etal (2005), and the one for the deDonder coordinates from Zhang (1986, eq. 3.26). In Fermi normal coordinates there are no time derivatives of the Riemann normal components or their spatial derivative, whereas in deDonder coordinates the lowest (second) order time derivative occurs in the third order correction coefficient. In affine null coordinates, however, we recognize a hierarchical order of the time derivative of the Riemann normal components and their spatial covariant derivatives\footnote{In deDonder coordinate, these time derivatives also occur in hierarchical order, \ie the $\ms{C}_{2k}$ correction coefficients of the metric along the geodesic depends on the $(2k)^{th}$ time derivative of the $k^{th}$ covariant derivative of $\mc{R}_{0i0j}.$}.  

\begin{table}
\caption{\label{tab:comp_deD_FNC_AN} 
Occurrence of the  contracted Riemann normal component $\ms{R}_{0i0j}n^in^j$, their contracted spatial covariant derivatives and their respective time derivative along the time-like geodesic in the metric component $g_{00}$ in affine null coordinates, Fermi normal coordinates and de Donder coordinates. The order  of the correction coefficient $\ms{C}_n$ corresponds to the power of $r^{n+1}$ in the radial expansion of $g_{00}$. \\} 
\begin{indented}
\item
\begin{tabular}{cccc}
\hline
order   of   & affine   & Fermi& deDonder\\[-3ex]
correction & null      & normal&  \\[-3ex]
coefficient & coordinates & coordinates&coordinates \\
\hline
$\ms{C}_1$ & $\mc{R}_{i0j0}n^in^j$ &$\mc{R}_{i0j0}n^in^j$&$\mc{R}_{i0j0}n^in^j$\\\hline
$\ms{C}_2$ & $\mc{R}_{i0j0;k}n^in^jn^{k}$ & $\mc{R}_{i0j0;k}n^in^jn^{k}$&$\mc{R}_{i0j0;k}n^in^jn^k$\\
 & $\partial(\mc{R}_{i0j0}n^in^j)/\partial\tau_w$ &&\\\hline
$\ms{C}_3$ & $\mc{R}_{i0j0;kl}n^in^jn^{k}n^{l}$ &$ \mc{R}_{i0j0;k l}n^in^jn^{k}n^{l}$&$\mc{R}_{i0j0;kl}n^in^jn^kn^l$\\
&$\partial(\mc{R}_{i0j0;k}n^in^jn^{k})/\partial\tau_w$ &&$\partial^2(\mc{R}_{i0j0}n^in^j) /\partial\tau_w^2$\\
&$\partial^2(\mc{R}_{i0j0}n^in^j)/\partial\tau_w^2$&&\\
\hline
\end{tabular} 

\end{indented}
\end{table}

Following the same line of arguments as for the time derivatives, one
can show that the third-order correction coefficients depend on the
square of the Riemann normal components. This behavior of the affine
null metric near the vertex was also observed by Poisson  and Vlasov
(2010), who derived an affine null metric for vacuum space-times.
Since the Bondi--Sachs metric and the affine null metric are related
by a transformation of the radial coordinate only, everything that we
said above about the affine null metric applies to the functions
$\ms{F}$, too.

As normal coordinates exist along $\bs{c}(\tau)$ and as a power series of a metric in normal coordinates can be transformed into a power series of a metric in Bondi-Sachs coordinates,
regularity implies the following restrictions on the power series expansion coefficients of $\ms{F}$ at the vertex:
\begin{itemize}
  \item[$(iv)$]  The correction coefficient $\ms{C}_n,(n>1)$ depends on the time derivatives of  order $(n-k)$ of the  correction coefficient $\mc{C}_k,\, (1\le k<n)$ and these time derivatives must be finite.

  \item[$(v)$] Non-linear coupling occurs at lowest order near the
    vertex in the third-order correction coefficient of $\ms{F}$, \ie
    at $\mc{O}(r^4)$ for $\gamma,\,\delta,\,\beta,\,\Phi$, and at
    $\mc{O}(r^3)$ for $U^A$.
\end{itemize}
The requirements $(i)-(v)$ state the general properties of the boundary conditions for  the Bondi--Sachs metric functions $\ms{F}$, when the vertex is a regular point in the null cone which is traced by a time-like geodesic. As these properties have been derived without the solution of the Einstein Equations (they can be considered as ``kinematical conditions''), they depend on the solution of  Einstein equations (which is a ``dynamical condition''). 

We summarize the general requirements on the boundary conditions of the functions $\ms{F}$ at $\ms{O}$ in the following \textbf{Vertex Lemma}:

\vspace{1ex}
\noindent 
{\it Let the Bondi--Sachs metric functions, $\ms{F}$,  be represented by power series expansions with respect to the areal distance coordinate $r$ at the vertex of a null cone on a time-like geodesic. 
    If the coefficients  of this radial expansion obey 
    \begin{enumerate}
  \item[(1)] the general properties $(i)-(v)$, 
  \item[(2)] and the Einstein equations, 
\end{enumerate} 
     then this power series expansion of $\ms{F}$ can be used to formulate boundary conditions for the metric functions $\ms{F}$ at the vertex of the null cone, and the vertex is a regular point in the null cone.  }

\vspace{1ex}
\noindent
In the next section, we explicitely calculate the boundary conditions for $\ms{F}$ at the vertex for
axisymmetric vacuum space-times.  We choose these space-times, to complete and extend the boundary conditions as used by Gomez \etal (1994) and Siebel \etal (2002), and to  recover and justify their {\it ad hoc} assumptions on the boundary condition employed in their numerical algorithms. In addition, axisymmetric space-times are the simplest\footnote{Spherically symmetric space-times are of no use here,
  because they possess no angular structure,\ie property (ii) of the
  vertex lemma cannot be demonstrated.} ones
 to show all the relevant addressed in the vertex lemma.
\section{Solutions for axially symmetric space-times}
\label{sec:sols}
A four-dimensional axially symmetric space-time $(\ms{M}^4,\,g)$ is a
space-time containing a time-like 2-surface $\ms{A}$ consisting of
points that are invariant under the action of a one-parametric, cyclic
group $G$ that is isomorphic to SO(2) (Carter, 1970, Stephani \etal
2003, and references therein). In particular, since the metric is
invariant under the action of $G$, \ie there exists a Killing vector
field $\bs{\xi}(\phi)$ in $(\ms{M}^4,\,g)$, where $\phi$ is the
parameter of the group $G$.  The vector field $\bs{\xi}(\phi)$ is
constant along the orbits of the group action, and the Lie derivative
$\ms{L}_{\bs{\xi}(\phi)} g_{\mu\nu}$ vanishes along the curves
generated by $\bs{\xi}(\phi)$. It can be shown (Carot, 2000, and
references therein) that the 2-surface $\ms{A}$, the so-called axis of
symmetry, is auto-parallel. Since $(\ms{M}^4,\,g)$ is endowed with a
metric, the fact that $\ms{A}$ is auto-parallel is equivalent to
$\ms{A}$ being totally geodesic (Spivak, 1999). As $\ms{A}$ is
time-like, the axis of symmetry contains a family of time-like
geodesics. Since these geodesics distinguish themselves from other
geodesics in $\ms{M}^4$ due to their axial symmetry, we call them {\it
  axial geodesics}. They are the natural choice to trace the origin
(vertex) of a Bondi--Sachs coordinate system in an axially symmetric
space-time.

Hereafter, let $\bs{c}(\tau)$ be an axial geodesic that is normalized
in such a way that $\tau$ is the proper time. Given an orthonormal
tetrad $\bs{e}_\mu(\tau)$ along $\bs{c}(\tau)$, we define a Fermi
normal coordinate system $y^\alpha$ along $\bs{c}(\tau)$, where
$\partial/\partial\tau$ is tangent to $\bs{c}(\tau)$ and $\tau$ is the
time coordinate. We further choose three mutually orthogonal
space-like vectors $\partial/\partial{y^i}$ at every point on
$\bs{c}(\tau)$, which are parallel transported along the axial
geodesic. Of these three vectors, the two vectors
$\partial/\partial{y^1}$ and $\partial/\partial{y^2}$ are normal to
the time-like 2-surface $\ms{A}$, while the vector
$\partial/\partial{y^3}$ is tangent to it. When $y^0=const$ and
$y^3=const$, the coordinates $y^1$ and $y^2$ label points of Killing
orbits of $G$ in $\ms{M}^4$ whose fixed-points are located on the
axial geodesic $\bs{c}(\tau)$. When instead $y^0=const$ and
$y^1=0=y^2$ holds, the coordinate $y^3$ labels points on the symmetry
axis $\ms{A}$.  In the Fermi normal coordinate system introduced above
the Killing vector has the form (Carot, 2000)
\begin{equation}
\label{eq:Kill_fermi}
\fl   \xi^\mr{a}(y^\mr{b}) = y^2\kron{\mr{a}}{1} -
                                         y^1\kron{\mr{a}}{2}\;\;.
\end{equation}

In the following, we will present two approaches to determine the
behavior of the Bondi--Sachs metric functions $\ms{F}$ near
$\bs{c}(\tau)$. In the first approach, in
Section\,\ref{sec:direct_fermi}, we solve the Einstein equations in
the Fermi normal coordinate system to obtain the Riemann normal
components, which are then used to calculate the Bondi--Sachs metric
functions from the expressions \eqref{eq:contractions_riem_A} --
\eqref{eq:contractions_riem_S}. In the second approach, in
Section.\,\ref{sec:power}, we expand the metric functions $\ms{F}$
into a power series obeying both axial symmetry and the limiting
behavior of $\ms{F}$ near $r=0$ as given in \eqref{eq:BS_F_gam} -
\eqref{eq:BS_F_Phi}. Subsequently, we solve the vacuum Einstein
equations in Bondi--Sachs coordinates for the power series
coefficients.
\subsection{Lowest order non-tivial boundary conditions  for the Bondi--Sachs metric derived directly
            from the Fermi metric}
\label{sec:direct_fermi}
The vacuum Einstein equations are $\mc{R}_{\mu\nu}=0$. In axial
symmetry, every tensor $T$ has to obey the Killing condition, \ie the
Lie derivative $\ms{L}_{\bs{\xi}(\phi)}T$ has to vanish. If we apply
this condition to the Riemann normal components $\mc{R}_{\mr{abcd}}$
in a Fermi normal coordinate system using the Killing vector
\eqref{eq:Kill_fermi}, we find the following relations among the
non-zero components of $\mc{R}_{\mr{abcd}}$:
\begin{eqnarray}
\label{eq:rime_axi}
\fl  &&\mc{R}_{0101}= 
 \mc{R}_{0202}\;\;,
 \mc{R}_{0113}=
 \mc{R}_{0223}\;\;,
 \mc{R}_{0312}=
 -2\mc{R}_{0123}\;\;,
 \mc{R}_{1313}=
 \mc{R}_{2323}\;,
 \mc{R}_{0303}\;,
 \mc{R}_{1212}.\nonumber\\
 \fl
\end{eqnarray}
For the further discussion we conveniently combine these non-zero
components into a set ${\mc{I} \in \{A,\,B,\,C,\,D,\,E,\,F\}}$ with
\begin{equation}
\label{ }
\fl 
A:=\mc{R}_{0101}\;,\;
B:=\mc{R}_{0303}\;,\;
C:=\mc{R}_{0113}\;,\;
D:=\mc{R}_{0123}\;,\;
E:=\mc{R}_{1212}\;,\;
F:=\mc{R}_{1313}\;.\;
\end{equation}
The  Ricci tensor reads in Fermi normal coordinates
\begin{equation}
\label{eq:exp_ricci}
\fl 
\mc{R}_{\alpha\beta}(y^\mr{a}) 
   =\left. \left(\begin{array}{cccc}-2A-B & 0 & 0 & -2C \\[-3ex]
                                        0 & A-\f{1}{2}(E+F) & 0 & 0 \\[-3ex]
                                        0 & 0 & A-\f{1}{2}(E+F) & 0 \\[-3ex]
                                      -2C & 0 & 0 & B-F 
                 \end{array}
    \right)\right |_{\bs{c}(\tau)} + \mc{O}(y^\mr{a})\;\;,
\end{equation}
where the functions $\mc{I}$ are evaluated along the axial geodesic
$\bs{c}(\tau)$.  We note that the function $D\big|_{\bs{c}(\tau)}$ is
not determined by the Ricci tensor. Hence, it is not determined by the
vacuum Einstein equations in Fermi normal coordinates, and thus can be
prescribed freely at the vertices of the null cones.  Solving the
vacuum Einstein equations for the zeroth-order coefficient of the
expansion \eqref{eq:exp_ricci} results in the solution
\begin{equation}
\label{ }
\fl 
B\big|_{\bs{c}(\tau)}= -2A\big|_{\bs{c}(\tau)}\;\;,\;\;
C\big|_{\bs{c}(\tau)}=0\;\;,\;\;
E\big|_{\bs{c}(\tau)}=4A\big|_{\bs{c}(\tau)}\;\;,\;\;
F\big|_{\bs{c}(\tau)} = -2A\big|_{\bs{c}(\tau)}\;\;,
\end{equation}
which depends only on the function $A\big|_{\bs{c}(\tau)}$ freely
specifiable along $\bs{c}(\tau)$.

Next we set ${A|_{\bs{c}(\tau)}(\tau_w)}:=
6\widetilde{\gamma}_2(\tau_w)$ and ${D|_{\bs{c}(\tau)}(\tau_w):=
  6\widetilde{\delta}_2(\tau_w)}$, and calculate the Bondi--Sachs
metric up to the first-order correction to the flat metric for an
axially symmetric vacuum space-time using
${\eqref{eq:contractions_riem_A} - \eqref{eq:contractions_riem_S}}$
and $\eqref{eq:BS_F_gam}- \eqref{eq:BS_F_Phi}$. We find
\numparts
\begin{eqnarray}
\fl  \gamma(x^\alpha)&=& r^2 \widetilde{\gamma}_2 (\tau_w) P^2_2(y)
+ \mc{O}(r^3)\;\;,
\label{eq:sol_fermi_vac_1st_gam}\\
\fl  \beta(x^\alpha)    &=& \mc{O}(r^3)\;\;,\label{eq:sol_fermi_vac_1st_beta}\\
\fl   U^\theta(x^\alpha) &=& -4 r  
    \widetilde{\gamma}_2(\tau_w) P^1_2(y) + \mc{O}(r^2)\;\;, \\
\fl  \Phi(x^\alpha)     &=& -6 r  
    \widetilde{\gamma}_2(\tau_w) P^0_2(y) + \mc{O}(r^3)\;\;,
\end{eqnarray} and 
\begin{eqnarray}
\fl   \delta(x^\alpha)&=& r^2   
    \widetilde{\delta}_2(\tau_w) P^2_2(y)+\mc{O}(r^3)\;\;,\\
\fl  U^\phi(x^\alpha)  &=& -4 r  
    \widetilde{\delta}_2(\tau_w)
    \bigg(\f{ P^1_2(y)}{\sin\theta}\bigg) +\mc{O}(r^2)\;\;,
\label{eq:sol_fermi_vac_1st_U31}
\end{eqnarray}
\endnumparts 
where $y:=-\cos\theta$ and $P^m_l(y)$ are the associated Legendre
polynomials of the first kind. The functions $\widetilde{\gamma}_2$
and $\widetilde{\delta}_2$ determine, respectively, the electric and
magnetic part of the Weyl tensor at the vertex. Moreover, they also
define the lowest order curvature contributions in the norm and twist
of the Killing vector w.r.t.\ their flat space values at the vertex,
respectively.

Equations \eqref{eq:sol_fermi_vac_1st_gam}-\eqref{eq:sol_fermi_vac_1st_U31} give the lowest non-trivial  order boundary conditions for $\ms{F}$ at a regular vertex for axially symmetric vacuum space-times.  Equation  \eqref{eq:sol_fermi_vac_1st_beta} shows that $\beta$ does not behave as $\mc{O}(r^2)$ as expected by property $(i)$ of the vertex lemma. This also demonstrates that if we had started with the expansions of Zhang (1986) to find the generic properties of the Bondi--Sachs metric at the vertex,  the most general limiting behavior of $\beta$ would be wrong. In fact it can be shown that for axially symmetric perfect fluid space-times $\beta$ is of $\mc{O}(r^2)$ at the vertex.

In principle, it is possible to calculate also the next order
correction coefficients with the above approach, but this involves
contractions of the covariant derivatives of the Riemann normal
components over five and more indices with the null vector $k^\mr{a}$
and its angular derivatives, \eg the $\mc{O}(r^3)$ coefficient of
$\gamma$ depends on $(\nabla_\mr{a}\mc{R}_{\mr{bcde}}) k^\mr{a}
k^\mr{b} k^\mr{c}_{,A} k^\mr{d} k^\mr{e}_{,B}$.  Thus, we describe in
section \ref{sec:power} another approach that avoids the tedious
calculation of these contractions, and also is easily extendable
to higher order.
\subsection{Regular boundary conditions for the 
            Bondi--Sachs metric derived from a power series 
            of $\ms{F}$ }
\label{sec:power}
To find axially symmetric solutions for the Bondi--Sachs metric
\eqref{eq:BS_trad}, we transform the Killing vector
\eqref{eq:Kill_fermi} to Bondi--Sachs coordinates $x^\alpha$, \ie
$\xi^\alpha(x^\alpha) = \kron{\alpha}{\phi}$. The Killing equations
$\ms{L}_{\bs{\xi}(\phi)}g_{\alpha\beta}(x^\mu)=0 $ imply that the
Bondi--Sachs metric functions $\ms{F}$ do not depend on the coordinate
$\phi$. To facilitate the computations, we employ the coordinate
transformation $y = -\cos\theta$ in the metric \eqref{eq:BS_trad}.
Points on the axis of symmetry, \ie the poles, are given by $y=\pm
1$. We further define the auxiliary function $s(y):=\sqrt{1-y^2}$, and
anticipating from \eqref{eq:BS_F_gam} - \eqref{eq:BS_F_Phi} we assume
the following expansions for the metric functions $\ms{F}$:
\numparts
\begin{eqnarray}
\fl  \gamma(\tau_w, r, y) & = &
    \sum_{n=1}^N (wr)^{n+1}\gamma_{n+1}(\tau_w,y) +
    \mc{O}\Big[(wr)^{N+2}\Big]\;\;,
\label{eq:ser_gamma}\\
\fl  \delta(\tau_w, r, y) & = &
    \sum_{n=1}^N (wr)^{n+1}\delta_{n+1}(\tau_w,y) + 
    \mc{O}\Big[(wr)^{N+2}\Big]\;\;,\\
\fl  \beta(\tau_w, r, y)  & = &
    \sum_{n=1}^N (wr)^{n+1}\beta_{n+1}(\tau_w,y) + 
    \mc{O}\Big[(wr)^{N+2}\Big]\;\;,\\
\fl  \Phi(\tau_w, r, y)   & = &\sum_{n=1}^N (wr)^{n+1}\Phi_{n+1}(\tau_w,y)+\mc{O}\Big[(wr)^{N+2}\Big]\;\;,\\
\fl  U^A(\tau_w, r, y)    & = &\sum_{n=1}^N (wr)^{n}U^A_{n}(\tau_w,y)+\mc{O}\Big[(wr)^{N+1}\Big]\;\;,\label{eq:ser_UA}
\end{eqnarray} 
\endnumparts
where the set of expansion coefficients $\ms{C}_n:= \{\gamma_{n+1},\,
\delta_{n+1},\, \beta_{n+1},\, \Phi_{n+1},\, U^A_n\}$ is calculated at
$r=0$, $N$ is the order up to which the field. Note, this expansion only respects property $(i)$ of the vertex lemma, and we do not impose other restrictions at this stage. After calculating the vacuum Einstein equations for the coefficients $\ms{C}_n$, we end up with coupled partial differential equations for the $\ms{C}_n$ with respect to $u$ and $y$. We solve these equations in general, and then restrict their solution to be regular, \ie finite, at the boundaries. This will provide us with the boundary conditions for the Bondi-Sachs metric at a regular vertex $\ms{O}$.  These regular boundary conditions at $\ms{O}$ will then also comply with the general properties stated in the vertex lemma.
In the following, we refer to $\ms{C}_n$ as
the $n^{th}-$order correction of the Bondi--Sachs metric functions
$\ms{F}$ with respect to a flat metric at $\ms{O}$ or simply the
$n^{th}-$order correction of $\ms{F}$.

In Section \ref{sec:vac_sol} and \ref{sec:lin_sol}, we solve the
vacuum Einstein equations, which can be grouped in Bondi--Sachs
coordinates into six so-called main equations, three supplementary
equations, and one trivial equation, respectively (Bondi \etal 1962,
Sachs, 1962, van der Burg, 1966, Winicour, 2012).  The six main
equation can be split further into two evolution equations for the
transverse-traceless part of the 2-metric $h_{AB}(\gamma,\,\delta)$,
and four hyper-surface equations for the variables $\beta,\,U^A,$ and
$\Phi$.  Furthermore, from the twice contracted Bianchi identities
follows the lemma (Bondi \etal 1962, Sachs, 1962, Tamburino  and
Winicour, 1966): {\it If the main equations hold on one null cone and
  if the optical expansion rate of the null rays does not vanish on
  this cone then the trivial equation is fulfilled algebraically and
  the supplementary equations hold if they are fulfilled at one radius
  $r$}. Consequently, we need to consider only the main equation to
find the solution of the Einstein equations at the vertex $\ms{O}$.
The supplementary equations then provide a check of this solution. The
Ricci tensor components and the quantities derived from it that appear
in the supplementary, hyper-surface and evolution equations
determining the corrections $\ms{C}_n$ are given in \ref{app:Einstein}
.
\subsubsection{Boundary conditions depending on $\ms{C}_1,\,\ms{C}_2$ and 
               $\ms{C}_3$} 
\label{sec:vac_sol}
According to property (v) of the vertex lemma and the expansion of a
metric in normal coordinates, see \eg relation
\eqref{eq:Riem_quad_fermi}, we expect the non-linear coupling of the
$\ms{C}_n$-coefficients to happen at lowest order in the
$\ms{C}_3$-coefficients. Indeed, using \eqref{eqapp:R_rr},
$\mc{R}_{rr}=0$ and omitting terms $\mc{O}(r^3)$ gives
\numparts
\begin{eqnarray}
\fl  \beta_2&=&0\;\;,\;\;
\label{eq:sol_vac_b2}\\
\fl  \beta_3&=&0\;\;,
\label{eq:sol_vac_b3}\\
\fl  \beta_4&=&\f{1}{2}\Big[(\gamma_2)^2+(\delta_2)^2\Big]\;\;.
\label{eq:sol_prelim_beta_4}
\end{eqnarray}
\endnumparts
Equation \eqref{eq:sol_prelim_beta_4} shows that the $\ms{C}_3$
correction of $\beta$ depends quadratically on the solution of the
$\ms{C}_1$ correction.

To find the solutions for the $\ms{C}_n,\,n\in\{1,2,3\}$, coefficients
we solve first for $\ms{C}_1$ and $\ms{C}_2$ setting the constant and
linear coefficients of the $r-$series in \eqref{eqapp:R_rr} -
\eqref{eqapp:R_delta} equal to zero. We next insert the solution of
the $\ms{C}_1$-correction into the quadratic coefficients of the
$r-$series of the Ricci tensor components in \eqref{eqapp:R_rr} -
\eqref{eqapp:R_delta} and solve for the $\ms{C}_3$-coefficient.

Inserting the solution $\beta_2=0 = \beta_3$ into \eqref{eqapp:R_ry} -
\eqref{eqapp:R_gamma} and utilizing $\mc{R}_{\alpha\beta} =0 $ we find
\numparts
\begin{eqnarray}
\fl   0 & =&
                    2U^y_1+2(s^2\gamma_2)_{,y} 
                     +\Big[5U^y_2+3(s^2\gamma_3)_{,y}  \Big]r w^{-1}                    
                     +\mc{O}(r^2)\;\;,\label{eq:vac_R_ry}\\
\fl   0&=&              
                    2U^\phi_1+2s^{-2}(s^2\delta_2)_{,y} 
                    +\Big[5U^\phi_2 +3s^{-2}(s^2\delta_3)_{,y}  \Big]r w^{-1}                    
                   +\mc{O}(r^2)\;\;,\label{eq:vac_R_rphi}\\
\fl   0  
  &=&
  -12\Phi_2  + 5U^y_{1,y} 
     +2s^{-2}\big(s^4\gamma_{2,y}\big)_{,y}-4\gamma_2    \nonumber\\
\fl     &&      
     +\Big[-16\Phi_3 +6U^y_{2,y}
               +2s^{-2}\big(s^4\gamma_{3,y}\big)_{,y}-4\gamma_3 \Big]wr
              +\mc{O}(r^2)  \;\;, \label{eq:vac_R_2d}\\
\fl  0
  &=&
        -12\gamma_2  +3s^{2}\big(s^{-2}U^y_1)_{,y}
        +\Big[12\gamma_{2,\tau_w}-24\gamma_3 
         +4s^{2}\big(s^{-2}U^y_2)_{,y}\Big]wr
          +\mc{O}(r^2)\;\;,\label{eq:vac_R_gamma}\\
\fl  0
        &=&
            12\delta_2-3s^2U^\phi_{1,y}    
           +\Big(-12\delta_{2,\tau_w}+24\delta_3-4s^2U^\phi_{2,y}\Big)wr
          +\mc{O}(r^2)\;\;.\label{eq:vac_R_delta}
\end{eqnarray}
\endnumparts
Combining algebraically the zeroth-order term in the $r-$expansion of
\eqref{eq:vac_R_ry} and \eqref{eq:vac_R_gamma} gives
\begin{equation}
\label{eq:dgl_g2}
\fl  0=s^2\gamma_{2,yy}-2y\gamma_{2,y}+
                    \Big(6-\f{4}{s^2}\Big)\gamma_2\;\;, 
\end{equation}
while the same procedure yields for the first-order terms 
\begin{equation}
\label{eq:dgl_g3}
\fl  0=s^2\gamma_{3,yy}-2y\gamma_{3,y}+
                    \Big(12-\f{4}{s^2}\Big)\gamma_3
                    -5\gamma_{2,\tau_w}\;\;, 
\end{equation}
and leads to similar equations as \eqref{eq:dgl_g2} and
\eqref{eq:dgl_g3} when applied to \eqref{eq:vac_R_rphi} and
\eqref{eq:vac_R_delta}, $\gamma_2$ being replaced by $\delta_2$, and
$\gamma_3$ by $\delta_3$, respectively.  Hence solving
\eqref{eq:dgl_g2} and \eqref{eq:dgl_g3} does not only give $\gamma_2$
and $\gamma_3$, but also provides the structure of a solution of
$\delta_2$ and $\delta_3$.  Inserting the ansatz
$\gamma_2(\tau,_w,\,y) = \widetilde{\gamma}_2(\tau_w)S(y)$ in
\eqref{eq:dgl_g2} yields
\begin{equation}
\label{eq:dgl_g2_decoup}
\fl  0=\widetilde{\gamma}_2(\tau_w)\bigg[(1-y^2)
   \f{d^2S}{dy^2}-2y\f{dS}{dy}+\Big(6-\f{4}{1-y^2}\Big)S\bigg]\;\;. 
\end{equation}
Since $\widetilde{\gamma}_2$ is an arbitrary function, the term
in the parenthesis has to vanish, which also provides us with an
associated Legendre differential equation for $S(y)$ of the form
\begin{equation}
\label{eq:alDGL}
\fl  0=(1-y^2)\f{d^2S}{dy^2}-2y\f{dS}{dy}+
                    \Big[l(l+1)-\f{m^2}{1-y^2}\Big]S\;\;, 
\end{equation}
with $l=2$ and $m=2$.  This equation holds for the Legendre
polynomials of the first kind, $P^m_l(y)$, and of the second kind,
$Q^m_l(y)$, respectively. Hence the most general solution of
\eqref{eq:alDGL} involves two integration constants $S_1$ and $S_2$,
\ie
\begin{displaymath}
\fl  S(y) = S_1 P^m_l(y)+S_2Q^m_l(y)\;\;. 
\end{displaymath}
Since the associated Legendre polynomials of the second kind,
$Q^m_l(y)$, are singular at the poles $y=\pm 1$, we set
$S_2=0$. Hereafter, we will always set integration constants connected
to Legendre polynomials of the second kind equal to zero. Thereby we
guarantee  regular solutions at the poles and point $(ii)$ of the vertex lemma, since the $P_l^m(y)$ are well defined  for all angles $y\in[-1, 1]$. A regular solution of
\eqref{eq:dgl_g2} assuming separation of variables is
\begin{equation}
\label{eq:sol_vac_g2}
\fl  \gamma_2(\tau_w,\,y) = 
                  \widetilde{\gamma}_2(\tau_w)P^2_2(y)\;\;,
\end{equation}
where we absorbed the integration constant $S_1$ into the arbitrary
function $\widetilde{\gamma}_2(\tau_w)$. To solve \eqref{eq:dgl_g3},
we make the ansatz
\begin{equation}
\label{eq:g3_ans}
\fl \gamma_3(\tau_w, y) = \sum_{k=2}^\infty\widetilde{\gamma}^l_3(\tau_w)P^2_l(y)\;\;,
\end{equation}
where the coefficients $\widetilde{\gamma}^l_3(\tau_w)$ depend on the
time at the vertex and are determined by the differential
equation \eqref{eq:dgl_g3}. Inserting \eqref{eq:g3_ans} into
\eqref{eq:dgl_g3}, reveals that $\widetilde{\gamma}_3(\tau_w):=\widetilde{\gamma}^3_3(\tau_w)$ is
freely specifiable function, $ \widetilde{\gamma}^2_3(\tau_w) =
5/6(d\widetilde{\gamma}_2/d\tau_w)$, and all remaining coefficients
$\widetilde{\gamma}^l_3(\tau_w)$ must be zero. Therefore, the solution
for $\gamma_3$ is
\begin{equation}
\label{eq:sol_vac_g3}
\fl  \gamma_3(\tau_w,y) = 
   \widetilde{\gamma}_3(\tau_w)P^2_3(y) +
   \f{5}{6}\Big[\f{d\widetilde{\gamma}_2}{d\tau_w}(\tau_w)\Big]
   P^2_2(y)\;\;. 
\end{equation}
\noindent
As already mentioned above, we can derive similar partial differential
equations as \eqref{eq:dgl_g2} and \eqref{eq:dgl_g3} for $\delta_2$
and $\delta_3$, \ie we may assume a solution for $\delta_2$ and
$\delta_3$ as
\begin{eqnarray}
\fl  \delta_2(\tau_w,y) &=& 
   \widetilde{\delta}_2(\tau_w)P^2_2(y)\;\;, 
\label{eq:sol_vac_d2}\\
\fl  \delta_3(\tau_w,y) &=& 
   \widetilde{\delta}_3(\tau_w)P^2_3(y) +
   \f{5}{6}\Big[\f{d\widetilde{\delta}_2}{d\tau_w}(\tau_w)\Big]
   P^2_2(y)\;\;,
\label{eq:sol_vac_d3} 
\end{eqnarray}
where both $\widetilde{\delta}_2 (\tau_w)$ and $\widetilde{\delta}_3
(\tau_w)$ are arbitrary functions of $\tau_w$.  

To solve for $U^A_1$ and $U^A_2$ we insert the solutions for
$\gamma_2$, $\gamma_3$, $\delta_2$, and $\delta_3$ into
\eqref{eq:vac_R_ry} and \eqref{eq:vac_R_rphi} and solve in the
constant and linear term of the $r-$series for the respective
variable. This yields
\begin{eqnarray}
\fl  U^y_1(\tau_w, y)   & = & 
   -4 \widetilde{\gamma}_2(\tau_w)P^1_2(y) s(y) \;\;,
\label{eq:sol_vac_U21}\\
\fl  U^\phi_1(\tau_w, y) & = & 
   -4 \widetilde{\delta}_2(\tau_w)  P^1_2(y)s^{-1}(y)\;\;,
\label{eq:sol_vac_U31}\\
\fl  U^y_2(\tau_w, y)   & = & 
   -\Big\{6\widetilde{\gamma}_3(\tau_w) P^1_3(y) + 
   2 \Big[\f{d\widetilde{\gamma}_2}{d\tau_w}(\tau_w)\Big]
   P^1_2\Big\}s(y)\;\;,
\label{eq:sol_vac_U22} \\
\fl  U^\phi_2(\tau_w, y) & = & 
   -\Big\{6\widetilde{\delta}_3(\tau_w) P^1_3(y) + 
   2 \Big[\f{d\widetilde{\delta}_2}{d\tau_w}(\tau_w)\Big]
   P^1_2\Big\}s^{-1}(y)\;\;.
\label{eq:sol_vac_U32}
\end{eqnarray}

The solutions for $\Phi_2$ and $\Phi_3$ are found inserting the
solution for $\gamma_2$, $\gamma_3$, $U^y_1$, and $U^y_2$ into
Eq.\,\eqref{eq:vac_R_2d} and solving the constant and linear term of
the $r-$series for $\Phi_2$ and $\Phi_3$, respectively. This gives
\begin{eqnarray}
\fl  \Phi_2 (\tau_w, y)& = & 
   -6\widetilde{\gamma}_2(\tau_w)P^0_2(y) \;\;,
\label{eq:sol_vac_Phi2}\\
\fl  \Phi_3(\tau_w, y) & = &
   -12\widetilde{\gamma}_3(\tau_w)P^0_3(y) -
   2\Big[\f{d\widetilde{\gamma}_2}{d\tau_w}(\tau_w)\Big]
   P^0_2(y)\;\;. 
\label{eq:sol_vac_Phi3}
\end{eqnarray}

Having obtained the complete solutions for the coefficients $\ms{C}_1$
in \eqref{eq:sol_vac_b2}, \eqref{eq:sol_vac_g2},
\eqref{eq:sol_vac_d2}, \eqref{eq:sol_vac_U21}, \eqref{eq:sol_vac_U31},
and \eqref{eq:sol_vac_Phi2}, and $\ms{C}_2$ in \eqref{eq:sol_vac_b3},
\eqref{eq:sol_vac_g3}, \eqref{eq:sol_vac_d3}, \eqref{eq:sol_vac_U22},
\eqref{eq:sol_vac_U32} and \eqref{eq:sol_vac_Phi3}, we now determine
the $\ms{C}_3$-coefficients of the metric function $\ms{F}$.  We
insert the solutions for $\gamma_2$ and $\delta_2$ into
Eq.\,\eqref{eq:sol_prelim_beta_4}, which gives
\begin{eqnarray}
\label{eq:sol_vac_b4}
\fl  \beta_4(\tau_w,\,y) &=& 
   \f{1}{2}\Big\{[\widetilde{\gamma}_2(\tau_w)]^2 +
   [\widetilde{\delta}_2(\tau_w)]^2\Big\}\Big[P^2_2(y)\Big]^2\;\;, 
\nonumber\\
\fl&=& \Big\{[\widetilde{\gamma}_2(\tau_w)]^2 + 
   [\widetilde{\delta}_2(\tau_w)]^2\Big\}
   \Big[\f{9}{7}P^2_2(y)-\f{3}{35}P^2_4(y)\Big] \;\;.
\end{eqnarray}
This equation shows that the solution can be expressed again as
a linear combination of associated Legendre polynomials.  We insert
the remaining $\ms{C}_2$ coefficients into {\eqref{eqapp:R_ry} -
\eqref{eqapp:R_delta}}, and then set these equations equal to zero.
Considering only the quadratic coefficients results in the following
equations:
\numparts
\begin{eqnarray}
\fl  0 & = &  9U^y_3 +
    4(s^2\gamma_4)_{,y} + 324ys^4\Big[(\gamma_2)^2+(\delta_2)^2\Big]\;\;,
\label{eq:u2_3_vac} \\
\fl  0 & = &  9U^\phi_3 + 4s^{-2}(s^2\delta_4)_{,y}\;\;,
\label{eq:u3_3_vac}\\
\fl  0 & = & -20\Phi_4  + 
    2s^{-2}\big(s^4\gamma_{4,y}\big)_{,y} -
    4\gamma_4+7U^y_{3,y} - 9(100y^4-32y^2+12)(\gamma_2)^2
\nonumber\\
\fl &&         +72(10y^4-11y^2+1)(\delta_2)^2 \;\;,
\label{eq:phi_4_vac}\\
\fl   0 & = &\!\!\! -40\gamma_4 + 
    5s^{2}\big(s^{-2}U^y_3\big)_{,y} + 40\gamma_{2,\tau_w\tau_w} + 
    240ys^2\gamma_{3,\tau_w} + 540s^4\Big[(\gamma_2)^2+(\delta_2)^2\Big] \nonumber\\
\fl    &&-
    720s^2\gamma_2^2\;,  \label{eq:g4_vac}\\
\fl   0 & = & 40\delta_4 - 5s^2U^\phi_{3,y} - 
    40s^2\delta_{2,\tau_w\tau_w} - 240ys^2\delta_{3,\tau_w} + 
    720s^2\delta_2\gamma_2\;\;. 
\label{eq:d4_vac}
\end{eqnarray}
\endnumparts 
Combining \eqref{eq:u2_3_vac} with \eqref{eq:g4_vac}, and
\eqref{eq:u3_3_vac} with \eqref{eq:d4_vac} yields
\begin{eqnarray}
\fl   0=s^2\gamma_{4,yy} - 2y\gamma_{4,y} +
   \Big(20-\f{4}{s^2}\Big)\gamma_4 - 
   18s^2\Big\{\gamma_{2,\tau_w\tau_w} + 6y\gamma_{3,\tau_w} - 
             9\big[(\gamma_2)^2-(\delta_2)^2]\Big\}\;\;,
\label{eq:dgl_g4_vac} \\
\fl   0=s^2\delta_{4,yy} - 2y\delta_{4,y} + 
    \Big(20-\f{4}{s^2}\Big)\delta_4 -
    18s^2\Big(\delta_{2,\tau_w\tau_w} + 6y\delta_{3,\tau_w} - 
              18\gamma_2\delta_2\Big)\;\;.
\label{eq:dgl_d4_vac}
\end{eqnarray}
Since the square of an associated Legendre polynomial can be expressed
by a linear combination of Legendre polynomials with the same degree
of $m$ but varying $l$, we assume that both $\gamma_4$ and $\delta_4$
obey an expansion in terms of $P^2_l(y)$ as
\begin{displaymath}
\fl   \gamma_4(\tau_w,y) = 
   \sum^\infty_{l=2}\widetilde{\gamma}_4^l(\tau_w)P^2_l(y)\;\;,\quad
                   \delta_4(\tau_w,y) = 
   \sum^\infty_{l=2}\widetilde{\delta}_4^l(\tau_w)P^2_l(y)\;\;.
\end{displaymath}
Inserting this ansatz into \eqref{eq:dgl_g4_vac} and
\eqref{eq:dgl_d4_vac} provides us with the solutions for the
coefficients $\widetilde{\gamma}_4^l$ and $\widetilde{\delta}_4^l$,
respectively,
\begin{eqnarray}
\fl  \gamma_4(\tau_w,y) & = & 
   \widetilde{\gamma}_4(\tau_w)P^2_4(y) + 
   \f{9}{10}\Big[\f{d\widetilde{\gamma}_3}{d\tau_w}(\tau_w)\Big]P^2_3(y)
\nonumber \\
\fl &&
   +\Bigg(\f{3}{7}\Big[\f{d^2\widetilde{\gamma}_2}{d\tau_w^2}
                        (\tau_w)\Big] - 
    \f{27}{7}\Big\{[\widetilde{\gamma}_2(\tau_w)]^2 -
                  [\widetilde{\delta}_2(\tau_w)]^2\Big\}\Bigg)
      P^2_2(y)\;\;,
\label{eq:sol_vac_g4}\\
\fl  \delta_4(\tau_w,y) & = & 
   \widetilde{\delta}_4(\tau_w)P^2_4(y) + 
   \f{9}{10}\Big[\f{d\widetilde{\delta}_3}{d\tau_w}(\tau_w)\Big]
   P^2_3(y)
\nonumber\\
\fl && 
   +\Bigg\{\f{3}{7}\Big[\f{d^2\widetilde{\delta}_2}{d\tau_w^2}
                        (\tau_w)\Big] -
           \f{54}{7}\widetilde{\gamma}_2(\tau_w)
                    \widetilde{\delta}_2(\tau_w)
    \Bigg\}P^2_2(y)\;\;,
\label{eq:sol_vac_d4}
\end{eqnarray}
where $\widetilde{\gamma}_4(\tau_w):=\widetilde{\gamma}^4_4(\tau_w)$ and $\widetilde{\delta}_4(\tau_w):=\widetilde{\delta}^4_4(\tau_w)$ are freely specifiable functions.
While  \eqref{eq:u2_3_vac}, \eqref{eq:u3_3_vac}, and
\eqref{eq:phi_4_vac} provide the solutions for $U^A_3$ and
$\Phi_4$, respectively:
\begin{eqnarray}
\fl  U^y_3(\tau_w,y) & = & -\bigg\{\Big[8\widetilde\gamma_4(\tau_w)+\f{72}{35}\widetilde{\gamma}_2^2(\tau)+\f{72}{35}\widetilde{\delta}_2^2(\tau)\Big]P^1_4(y)
                                                         +4\Big[\f{d\widetilde{\gamma}_3}{d\tau_w}(\tau_w)\Big]P^1_3(y) \nonumber\\
\fl&&+\Big\{\f{16}{21}\f{d^2\widetilde{\gamma}_2}{d\tau_w^2}(\tau_w)-\f{96}{7}[\widetilde{\gamma}_2(\tau_w)]^2\Big\}P^1_2(y)                                                          \bigg\}s(y) \;\;,\\
\fl   U^\phi_3(\tau_w,y) & = & -\bigg\{8\widetilde{\delta}_4(\tau_w)P^1_4(y)
                                                         +4\Big[\f{d\widetilde{\delta}_3}{d\tau_w}(\tau_w)\Big]P^1_3(y) \nonumber\\
\fl&&\quad+\Big[\f{16}{21}\f{d^2\widetilde{\delta}_2}{d\tau_w^2}(\tau_w)-\f{96}{7}\widetilde{\gamma}_2(\tau_w)\widetilde{\delta}_2(\tau_w)\Big]P^1_2(y)                                                          \bigg\}s^{-1}(y)\;\;, \\
\fl  \Phi_4(\tau_w,y)&=& -\bigg\{20\widetilde{\gamma}_4(\tau_w)+\f{864}{35}[\widetilde{\gamma}_2(\tau_w)]^2+\f{216}{35}[\widetilde{\delta}_2(\tau_w)]^2\bigg\}P^0_4(y)
     -6\Big[\f{d\widetilde{\gamma}_3}{d\tau_w}(\tau_w)\Big]P^0_3(y)\nonumber\\
\fl &&
        -\bigg(\f{4}{7}\f{d^2\widetilde{\gamma}_2}{d\tau_w^2}(\tau_w) -\f{24}{7}\Big\{[\widetilde{\gamma}_2(\tau_w)]^2+[\widetilde{\delta}_2(\tau_w)]^2\Big\}\bigg)P^0_2(y)
        \nonumber\\
\fl &&        
        -\f{48}{5}[\widetilde{\gamma}_2(\tau_w)]^2-\f{12}{5}[\widetilde{\delta}_2(\tau_w)]^2\;\;.\label{eq:sol_vac_Phi4}
\end{eqnarray}

Equations\,\eqref{eq:sol_vac_b4}, \eqref{eq:sol_vac_g4} -
\eqref{eq:sol_vac_Phi4} constitute a solution for the
$\ms{C}_3-$coefficients of $\ms{F}$. 
The solution  for the corrections $\ms{C}_2$ and $\ms{C}_3$  demonstrates points $(iii),\,(iv)$ and $(v)$ of the vertex lemma, as the expansion coefficients contain specific numerical factors (point $(iii)$). Time derivatives (point $(iv)$) of the lower order coefficients $\ms{C}_1$ and $\ms{C}_2$ occur hierachically ordered, \ie $\ms{C}_2$ depends on the first time derivative of $\ms{C}_1$ whereas $\ms{C}_3$ depends on the second time derivative of $\ms{C}_1$ and on the first time derivative of $\ms{C}_2$. The non-linear coupling (point $(v)$) in the correction coefficients occurs at lowest order in $\ms{C}_3$, since $\ms{C}_3$ depends on $(\ms{C}_1)^2$. If one  uses the solution for $\ms{C}_n,n=(1,2,3)$ as boundary conditions for the Bondi--Sachs metric in numerical simulations for axially symmetric space, the free functions $\widetilde{\gamma}_2$ and  $\widetilde{\delta}_2$ must be at least twice differentiable with finite derivatives, $\widetilde{\gamma}_3$ and  $\widetilde{\delta}_3$ must be differentiable with finite derivatives,  whereas  $\widetilde{\gamma}_4$ and  $\widetilde{\delta}_4$ must be continuous. Thus, requiring the vertex to be a regular point in the null cone rigidly fixes the boundary conditions for the metric at the vertex. The only freedom left is the choice of the functions $\widetilde{\gamma}_n$ and $\widetilde{\delta}_n,\,n\in\{1,2,3\}.$  
\subsubsection{Boundary conditions that are linear in $\ms{C}_n$ and 
               valid for $n\ge 1$}
\label{sec:lin_sol}
To investigate the boundary conditions of $\ms{F}$
further, we consider \eqref{eq:ser_gamma} - \eqref{eq:ser_UA} for an
arbitrary value of $N$. Since the metric functions $\ms{F}$ are power
series in $r$ with coefficients $\ms{C}_n$, the corresponding Ricci
tensor can be written as
\begin{displaymath}
\fl  \mc{R}_{\mu\nu}(\tau_w,\,r,\,y) = 
   \sum_{n=1}^N r^n \mc{R}^{(n)}_{\mu\nu}(\tau_w,\,y)\;\;,
\end{displaymath}
\ie as a power series in $r$, too.
As the coefficients $\mc{R}^{(n)}_{\mu\nu}$ will, in general, not
depend linearly on $\ms{C}_n$, we linearize them with respect to
$\ms{C}_n$ to firstly simplify the calculations,  and secondly to find the qualitative dependence of the hierarchical order of the time derivatives addressed in property $(v)$ of the vertex lemma.
 The linearized Ricci tensor
components for the main equations of the vacuum Einstein equations are
given in \ref{app:lin}. Introducing the operator
\begin{equation}
\label{ }
\fl  \mathbb{L}_l(f):=
   (1-y^2)f_{,yy}-2yf_{,y}+\Big[l(l+1)-\f{4}{1-y^2}\Big]f
\end{equation}
for a function $f$ that depends on $y$, allows us to write some of the
upcoming equations in a more compact way. In particular, if
$f(y)=P^2_l(y)$, we find
\begin{equation}
\label{eq:relation_P2_l}
\fl   \mathbb{L}_l\Big[P^2_j(y)\Big] = 
    b^j_{\p{j}l} P^2_j(y) \;\;,\qquad b^j_{\p{j}l}:=l(l+1)-j(j+1)\;\;,
\end{equation}
where there is no summation performed over $j$ on the right hand side
of the first expression. Note that the associated Legendre polynomial
$P^2_l(y)$ commutes
\footnote{A similar relation as \eqref{eq:relation_P2_l} could be
  derived for the Legendre polynomials of the second kind, $Q^2_l(y)$,
  but we do not investigate this further here as these polynomials are
  not defined at $y=\pm1$.}
with the operator $\mathbb{L}_l$, because $b^l_{\p{l}l}=0$.

From \eqref{eq:Rrr_lin} and the vacuum Einstein equation
$\mc{R}^{(n)}_{rr}=0$, we deduce
\begin{equation}
\label{eq:solbeta_np1}
\fl  \beta_{n+1}(\tau_w,\,r,\,y)=0\;\;, 
   \qquad \forall \;n\ge 1\;\;. 
\end{equation}
Inserting the solution for $\beta_n$ into \eqref{eq:Rr2_lin} -
\eqref{eq:Rd_lin_n} and combining $\mc{R}^{(1)}_{rA}=0$ from
\eqref{eq:Rr2_lin} and \eqref{eq:Rr3_lin} with
$\mc{R}^{(2)}_{(\gamma)}=0$ and $\mc{R}^{(2)}_{(\delta)}=0$ from
\eqref{eq:Rg_lin_2} and \eqref{eq:Rd_lin_2}, respectively, yields
\begin{equation}
\label{ }
\fl   0 = \mathbb{L}_2(\gamma_2)\;\;,\;\;
                   0 = \mathbb{L}_2(\delta_2)\;\;. 
\label{eq:lin_g2_d2}\\
\end{equation} 

Since the associated Legendre polynomial $P^2_2(y)$ commutes with
$\mathbb{L}_2$, regular solutions of \eqref{eq:lin_g2_d2} for
$y\in[-1,\,1]$ may be assumed to have the form
\begin{equation}
\label{ }
\fl   
    \gamma_2(\tau_w,\,y) = \widetilde{\gamma}^2_2(\tau_w)P^2_2(y)\;\;\;,
    \quad
    \delta_2(\tau_w,\,y) = \widetilde{\delta}^2_2(\tau_w)P^2_2(y)\;\;.
\end{equation}
where $\widetilde{\gamma}^2_2(\tau_w)$ and $\widetilde{\delta}^2_2
(\tau_w)$ are arbitrary functions depending on the time $\tau_w$ along
the geodesic $\bs{c}(\tau)$. Given the solutions for $\gamma_2$ and
$\delta_2$, and from $\mc{R}^{(1)}_{rA}=0$, we derive
\begin{equation}
\label{ }
\fl  
   U^y_1  (\tau_w,\,y) = -4 \widetilde{\gamma}^2_2(\tau_w)
                        P^1_2(y)s(y)\;\;,
\qquad
   U^\phi_1(\tau_w,\,y) = -4 \widetilde{\delta}^2_2(\tau_w)
                        P^1_2(y)s^{-1}(y)\;\;.
\end{equation} 
Setting $\mc{R}^{(0)}_{(2D)}$ in \eqref{eq:R2d_lin} equal to zero and
inserting $\gamma_2$ and $U^y_1$ leads to
\begin{equation}
\label{ }
\fl  \Phi_2(\tau_w,\,y) = -6\widetilde{\gamma}^2_2
                                      (\tau_w)P^0_2(y)\;\;,
\end{equation}
which shows that the solution for the lowest order correction
$\ms{C}_1$ obtained with the linear approach is equivalent to that of
the generic approach in the previous section.

To find the solutions for $\ms{C}_n$, $n>1$, we assume hereafter
$n>1$.  Combining $\mc{R}^{(n)}_{rA}=0$ from Eqs.\,\eqref{eq:Rr2_lin}
and \eqref{eq:Rr3_lin} with $\mc{R}^{(n+1)}_{(\gamma)}=0$ and
$\mc{R}^{(n+1)}_{(\delta)}=0$ from Eqs.\,\eqref{eq:Rg_lin_n} and
\eqref{eq:Rd_lin_n}, respectively, gives rise to two equations that
determine the solutions for $\gamma_{n+1}$ and $\delta_{n+1}$:
\begin{eqnarray}
\label{ }
\fl  0&=&\mathbb{L}_{n+1} (\gamma_{n+1}) + 
                      a_{n+1} \gamma_{n,\tau_w}\;\;,\qquad 
                  0 = \mathbb{L}_{n+1} (\delta_{n+1}) + 
                      a_{n+1} \delta_{n,\tau_w}\;\;,\qquad
\label{eq:gn_dn}
\end{eqnarray}
where
\begin{equation}
\label{ }
\fl  a_n:=-\f{2(n+2)(n-1)}{(n+1)}\;\;.
\end{equation}
The expressions in \eqref{eq:gn_dn} show that the solutions for
$\gamma_n$ and $\delta_n$ obey the same type of equation, \ie knowing
one of the solutions gives us the form of the other solution, too.

We treat the solutions for $\gamma_n$ and $\delta_n$ together by
defining $I^A_k:= \gamma_k\kron{A}{y} + \delta_k\kron{A}{\phi}$ with
$k\ge2$, whereby \eqref{eq:gn_dn} becomes
\begin{equation}
\label{eq:Ia}
\fl  0=\mathbb{L}_{n+1} (I^A_{n+1}) + a_{n+1} I^A_{n,\tau_w}\;\;.
\end{equation}
Making  the ansatz
\begin{equation}
\label{eq:ansI_np1}
\fl  I^A_{n+1}(\tau_w, y)=
   \sum^{n+1}_{k=2}\widetilde{I}^A_{n+1.k}(\tau_w)P^2_k(y)\;\;,
\end{equation}
which we insert into \eqref{eq:Ia} and obtain for $n=2,3$, and $4$
\numparts
\begin{eqnarray}
\fl  0 =  \mathbb{L}_3({I}^A_3) + a_3{I}^A_{2,\tau_w} \!
                   &=&   b^3_{\p{3}3}\widetilde{I}^A_{3.3} P^2_3 
                       + \bigg(b^2_{\p{2}3}\widetilde{I}^A_{3.2} 
                       + a_3\f{d\widetilde{I}^A_{2.2}}{d\tau_w}
                         \bigg) P^2_2\;\;,
\label{eq:cond_IA_3}\\
\fl  0 =  \mathbb{L}_4(I^A_4)+a_4{I}^A_{3,\tau_w} \!
    &=& b^4_{\p{4}4}\widetilde{I}^A_{4.4} P^2_4 
    +\! \bigg(\!b^3_{\p{3}4}\widetilde{I}^A_{4.3}  
    + a_4\f{d{\widetilde{I}}^A_{3.3}}{d\tau_w}\bigg)P^2_3
    +\! \bigg(\!b^2_{\p{2}4}\widetilde{I}^A_{4.2} 
    + a_4\f{d\widetilde{I}^A_{3.2}}{d\tau_w}\bigg)P^2_2\;,
\nonumber\\
    \fl&&\\
\fl  0 = \mathbb{L}_5(I^A_5)+a_5{I}^A_{4,\tau_w}\! 
    &=& b^5_{\p{5}5}\widetilde{I}^A_{5.5} P^2_5 
           +\! \bigg(\!b^4_{\p{4}5}\widetilde{I}^A_{5.4} 
           + a_5\f{d\widetilde{I}^A_{4.4}}{d\tau_w} \bigg)P^2_4    
           +\! \bigg(\!b^3_{\p{3}5}\widetilde{I}^A_{5.3} 
           + a_5\f{d\widetilde{I}^A_{4.3}}{d\tau_w} \bigg)P^2_3 
\nonumber \\
\fl       &&    
           + \bigg(b^2_{\p{2}5}\widetilde{I}^A_{5.2} +
           a_5\f{d\widetilde{I}^A_{4.2}}{d\tau_w}\bigg)P^2_2\;\;.  
\label{eq:cond_IA_5}  
\end{eqnarray}
\endnumparts

Since the associated Legendre polynomials depend on the arbitrary
angle $y$ and are non-zero in general, the coefficients of the
spectral series in terms of $P^2_l$ must vanish in order to fulfill
\eqref{eq:cond_IA_3} - \eqref{eq:cond_IA_5}. Since the diagonal
coefficients $b^l_{\p{l}l}$ are equal to zero, the functions
$\widetilde{I}^A_{l.l}(\tau_w)$ can be chosen arbitrarily. For the
other functions, we find
\numparts
 \begin{eqnarray}
\fl  \widetilde{I}^A_{3.2}& =& 
  \!\! - \bigg[\f{a_3}{b^2_{\p{2}3}}\Bigg]\f{d}{d\tau_w}\widetilde{I}^A_{2.2}\;\;,
\label{eq:I32}\\
\fl  \widetilde{I}^A_{4.2}& =& \!\!\quad \;
      \bigg[\f{a_3}{b^2_{\p{2}3}}\f{a_4}{b^2_{\p{2}4}}\bigg]\f{d ^2}{d\tau_w^2}
      \widetilde{I}^A_{2.2}\;\;,\quad\quad\!\!
   \widetilde{I}^A_{4.3} = -\bigg[\f{a_4}{b^3_{\p{2}4}}\bigg]\f{d }{d\tau_w}
                          \widetilde{I}^A_{3.3}\;\;
\label{eq:I42},\\
\fl  
   \widetilde{I}^A_{5.2}&=& \!\!-\bigg[\f{a_3}{b^2_{\p{2}3}}\f{a_4}{b^2_{\p{2}4}} 
     \f{a_5}{b^2_{\p{2}5}} \bigg]\f{d^3}{d\tau_w^3}\widetilde{I}^A_{2.2}\;, \;
   \widetilde{I}^A_{5.3} =   \bigg[\f{a_4}{b^3_{\p{2}4}}\f{a_5}{b^3_{\p{2}5}}\bigg]
   \f{d^2 }{d\tau_w^2}\widetilde{I}^A_{3.3}\;,\,
   \widetilde{I}^A_{5.4} = -\bigg[\f{a_5}{b^4_{\p{2}5}}\bigg]\f{d }{d\tau_w}
                           \widetilde{I}^A_{4.4}\;.
\label{eq:I52}
\end{eqnarray}
\endnumparts

Defining $J^A_n(\tau_w):= \widetilde{\gamma}_n(\tau_w)\kron{A}{y} +
\widetilde{\delta}_n(\tau_w)\kron{A}{\phi}$ with $n \ge 2$, where
$\widetilde{\gamma}_n(\tau_w)$ and $\widetilde{\delta}_n(\tau_w)$ are
arbitrary functions of $\tau_w$, we deduce from the recursive
behavior of $\widetilde{I}^A_{n.k}(\tau_w)$ in \eqref{eq:I32} -
\eqref{eq:I52} the general form of the time dependent coefficients of
$I^A_{n+1}(\tau_w, y)$ as
\begin{equation}
\label{eq:solcoeffI_np1}
\fl  
   \widetilde{I}^A_{n+1.l}(\tau_w) = 
   \cases{J^A_{n+1}(\tau_w) & for $ l=(n+1)$\;\;,\\
               c^{n+1}_{\p{n+1}l}\bigg[\f{d^{n-l}}{d\tau_w^{n-l}}J^A_l(\tau_w) 
                     \bigg]& for $ 2\le l\le n$\;\;,\\}
\end{equation}
where 
\begin{equation}
\label{eq:c_nk}
\fl   c^n_{\p{n}k}:=
    (-1)^{n+k}\f{\prod^n_{s=3} a_s 
       \prod^k_{t=3} b^k_{\p{k}t}}{\prod^k_{s=3} a_s \prod^n_{t=3} b^k_{\p{k}t}} 
    = \f{\prod^{n}_{s=k+1} \Big[\f{2 (s+2)(s-1)}{s+1}\Big]}{
         \prod^n_{t=k+1}\Big[t(t+1)-k(k+1)\Big]}\;\;. 
\end{equation}
The last term in the right hand side of \eqref{eq:c_nk} can be expressed by factorials using the computer algebra program Maple and some properties\footnote{Denoting with $\Gamma(n)$ the Gamma function, we use $\Gamma(n+1) =n!$ and $\Gamma(n+1/2)=2^{-2n}\pi^{1/2}(2n)!/n!$.} of the Gamma function, 
\begin{equation}
\label{eq:C_nk_fin}
\fl c^n_{\p{n}k} = \f{2^{n-k-1} k(n+2)(2k+2)!(n-1)!}{(k+2)!(n+k+1)!(n-k)!}\;\;.
\end{equation}
 
We find the solution for $U^A_n$ with $n>1$ by inserting $I^A_n$ into
$R^{(n)}_{rA}=0$, which leads to
\begin{eqnarray}
\label{eq:solU_n}
\fl U^A_n(\tau_w,y) &=& -\f{2(n+1)}{n(n+3)}
   \sum^{n+1}_{l=2}\widetilde{I}^A_{n+1.l}(\tau_w)
   \Big[(l+2)(l-1)\Big]P^1_l(y)q^A(y)\;\;,
\end{eqnarray}
where there is no summation performed over the index $A$ on the right
hand side, and $q^A:= s\kron{A}{y} + s^{-1}\kron{A}{\phi}$.  Setting
$\mc{R}^{(n)}_{(2D)}$ in \eqref{eq:R2d_lin} for $n>1$ equal to zero
and inserting $I^y_{n+1}$ and $U^y_n$, gives rise to the solution (for
$n>1$)
\begin{equation}
\label{eq:solPhi_np1}
\fl \Phi_{n+1}(\tau_w,\,y) =  - \sum^{n+1}_{l=2} 
    \Bigg[\f{l(l+1)(l+2)(l-1)}{n(n+3)}\Bigg] 
    \widetilde{I}^y_{n+1.l}(\tau_w) P^0_l(y) \;\;.                                                                 \\
\end{equation}

In summary, the axially symmetric vacuum solution for $\ms{C}_n$ with
arbitrary $n>1$ is given by $\beta_{n+1}$ in \eqref{eq:solbeta_np1},
the coefficients $\widetilde{I}_{n+1.l}$ in \eqref{eq:solcoeffI_np1}
for the power series \eqref{eq:ansI_np1}, $U^A_n$ in
\eqref{eq:solU_n}, and $\Phi_{n+1}$ in \eqref{eq:solPhi_np1},
respectively.  For this solution each set of coefficients $\ms{C}_n$
is determined by two functions $\widetilde{\gamma}_{n+1}(\tau_w)$ and
$\widetilde{\delta}_{n+1}(\tau_w)$, which can be prescribed freely
along the geodesic $\bs{c}(\tau)$.
\section{Summary and discussion}\label{sec:SumDiss}
We studied the boundary conditions of the Bondi--Sachs metric functions  $\ms{F}\in \{\gamma,\, \delta,\, \beta,\,
U^A,\, \Phi\}$ at the vertices of null cones
assuming that these vertices are traced by a time-like geodesic
$\bs{c}(\tau)$ and that the observer emitting the null rays from this
geodesic is an inertial observer.  General requirements of these boundary conditions were found in three major calculational steps after assuming $\bs{c}(\tau)$ was contained in a convex normal neighborhood:  In the first step, we constructed a metric
in Fermi normal coordinates along $\bs{c}(\tau)$. In the second step, we defined affine null
coordinates at $\bs{c}(\tau)${, where the radial coordinate is an affine parameter,} and 
transformed the metric  from Fermi normal
coordinates to an affine null 
metric, which agrees with that of Ellis \etal (1985), Poisson (2004, 2005)  and Poisson \etal (2006a, 2006b, 2010, 2011). 
In the third step, we calculated a Bondi--Sachs metric along
$\bs{c}(\tau)$,  by changing in the affine null metric the radial coordinate over to an areal distance coordinate $r$ that defines the radial coordinate in the Bondi--Sachs metric. 
The boundary conditions of  $\ms{F}$ at the vertex were found as Taylor series in terms of $r$ with respect to  a flat metric along $\bs{c}(\tau)$ where the expansion coefficients of the series are called correction coefficients $\ms{C}_n$.

Regularity at the vertex  implied the following five general requirements on the power series of $\ms{F}$:

\begin{enumerate}
\item[$(i)$]    The power series of $\ms{F}$  must
    start at $r=0$ with a certain positive power of $r$, in general  $\gamma,\delta,\,\beta,\,\Phi$ are of $\mc{O}(r^2)$ and $U^A$ of $\mc{O}(r)$.\footnote{If the curve tracing the vertices is no geodesic but a
  general time-like curve, with a tangent vector $u^\alpha$, the
  lowest order terms  depend differently on $r$, and one has to consider the acceleration, $a^\mu:=u^\alpha
  \nabla_\alpha u^\mu$, of the curve. Poisson \etal (2011)
  consider an affine null metric along such a general time-like
  curve. Transforming this metric to a Bondi--Sachs metric tells
    one how property $(i)$ should be changed in that
    case. However, with our assumption of axial symmetry, it is
  natural to assume that the curve is a geodesic on the axis of
  symmetry.}.

\item[$(ii)$] 
    The  coefficients $\ms{C}_n$  have a rigid angular structure that is given by  polynomials of harmonic and regular functions on  2-spheres centered on the vertices.

\item[$(iii)$] The polynomial coefficients of the harmonic base functions of $\ms{C}_n$  carry
  strict numerical factors depending on the physical problem under
  consideration.
  
  \item[$(iv)$] The  correction coefficient $\ms{C}_n,(n>1)$ depends on the  time derivatives of  order $(n-k)$ of the correction coefficient $\mc{C}_k,\, (1\le k<n)$  and these time derivatives must be finite. 
  
  \item[$(v)$] Non-linear coupling occurs at lowest order near the
    vertex in the third-order correction coefficient of $\ms{F}$, \ie
    at $\mc{O}(r^4)$ for $\gamma,\,\delta,\,\beta,\,\Phi$, and at
    $\mc{O}(r^3)$ for $U^A$.
\end{enumerate}
Requirements $(i)-(v)$ on the boundary conditions of $\ms{F}$ are kinematical regularity conditions, as they are derived from a regular metric along the geodesic tracing the vertices. The Einstein equations are not used in this establishing these conditions. The requirement that the $\ms{C}_n$ have to obey Einsteins equations, this is a dynamical condition.

We summarized these general requirements on the boundary conditions of $\ms{F}$ in a  \textbf{Vertex Lemma},  which also
answers question (I) from the introduction:

\vspace{1ex}
\noindent 
{\it Let the Bondi--Sachs metric functions, $\ms{F}$,  be represented by power series expansions with respect to the areal distance coordinate $r$ at the vertex of a null cone on a time-like geodesic. 
    If the coefficients  of this radial expansion obey 
    \begin{enumerate}
  \item[(1)] the general properties $(i)-(v)$, 
  \item[(2)] and the Einstein equations, 
\end{enumerate} 
     then this power series expansion of $\ms{F}$ can be used to formulate boundary conditions for the metric functions $\ms{F}$ at the vertex of the null cone, and the vertex is a regular point in the null cone.  }    

\vspace{1ex}
\noindent
We investigated the implication of the vertex lemma for axially axisymmetric vacuum space-times, which
are the simplest space-times that allow us to demonstrate its relevant features. We also corrected and generalized the boundary conditions for such space times existing in the literature (Gomez \etal, 1994, Siebel, \etal, 2001). Since axially symmetric space-times contain a
designated class of observers defined by time-like geodesics on the
axis of symmetry, we introduce a Fermi normal coordinate system with
respect to one of these axial geodesics $\bs{c}(\tau)$.

We then proposed two approaches to answer question (II) of the
introduction, \ie how one can determine the boundary conditions of the
Bondi--Sachs metric at the vertex.
  
In the first one, we calculated the Riemann normal components in the
Fermi normal coordinate system, and contracted these components with a
null vector at $\bs{c}(\tau)$, its angular derivatives, and the
tangent vector of $\bs{c}(\tau)$. These contracted Riemann normal
component provided first-order correction coefficients of the
Bondi--Sachs metric with respect to a flat metric at the
vertex. Although this approach works well for
the first-order corrections, it is unsuited to find the higher order
correction coefficients, because one would have to perform tedious
full contractions of covariant derivatives of the Riemann normal
components, and more importantly, because the approach itself cannot
easily be extended to higher order.

In the second approach we proposed the usage of a power series
expansion of the Bondi--Sachs variables $\ms{F}$, where the functions
are expanded with respect to positive powers $r$, the coefficients
depending on the time at the vertex and the direction angle of the
emanating null ray. The lowest powers of $r$ in these series were
chosen to agree with those given in property $(i)$ of the vertex
lemma. We then showed
that the other properties of the vertex lemma $(ii-v)$ follow when
integrating the Einstein equations, considering only the $\ms{C}_1-$,
$\ms{C}_2-$, and $\ms{C}_3-$correction coefficients.
  
Considering  property $(ii)$ of the vertex lemma, we solved the partial differential equations for the $\ms{C}_n$, and that they can in principle depend on the associated Legendre polynomials of first and second kind, $P^m_l(y)$ and $Q^l_m(y)$, respectively. Since the $Q^m_l(y)$ are singular at the poles, $y=\pm 1$, we rejected them as possible solution. This is in agreement with the `near-roundness'  condition of Choquet-Bruhat \etal (2010). The polynomials $P^m_l(y)$ are the harmonic and regular functions mentioned in property $(ii)$ of the vertex lemma.
We further note that the polynomials,
$Q^m_l(y)$, were not encountered in the first approach where the
Bondi--Sachs metric was determined directly from the Fermi metric,
because the regularity at the poles is already assured by choosing the
coordinate frame to be regular, \ie choosing a Fermi normal coordinate
system along $\bs{c}(\tau)$.
  
We also confirmed properties $(iii)$ and $(iv)$ of the vertex lemma, by
exploiting the fact that the spectral expansion of $\ms{C}_n,\,n>1,$
contains time derivatives of the indicated hierarchy (see table \ref{tab:timedriv_hierachy}) and 
carries numerical factors  following from the solution of the
Einstein equations.
  
Finally, we proved property $(v)$ of the vertex lemma by showing that
the solution for the $\ms{C}_3$ coefficient depends linearly on the
square of $\ms{C}_1$.

For an axially symmetric vacuum space-time the solutions for each of
the $\ms{C}_1-$, $\ms{C}_2-$, and $\ms{C}_3-$coefficients
depend on two time-dependent
functions along $\bs{c}(\tau)$ which can be prescribed freely,
\ie both functions are not constrained by the vacuum Einstein
equations. If these two free functions are $(3-n)$ times
differentiable with finite derivatives along $\bs{c}(\tau)$, the
solutions of $\ms{C}_1,\,\ms{C}_2$ and $\ms{C}_3$ provide the explicit
boundary conditions for the Bondi--Sachs metric with a regular vertex
along $\bs{c}(\tau)$ and for an axially symmetric vacuum. These six free functions correspond to the $l=2,3,4$- multipoles of the electric and magnetic part of the Weyl tensor evaluated at the vertex in agreement with the findings of Thorne(1980), Hartle \& Thorne (1985) and Zhang (1986), who found the metric expansions of a vacuum space-time along a timeline geodesic in deDonder coordinates.
The boundary conditions found in the second approach approximate a regular vertex up to third-order
corrections in Fermi normal coordinates and incorporate the non-linear
coupling due to curvature at lowest order. Thus, they provide the
answer to question (III) raised in the introduction.

As an example for such non-linear boundary conditions let us consider
the following six functions
\begin{equation}
\label{eq:free_SIMPLE}
\fl  \widetilde{\gamma}_2(\tau_w) = K\;\;,\;\;
\widetilde{\gamma}_3(\tau_w) = 0\;\;,\;\;
\widetilde{\gamma}_4(\tau_w)  = \f{9}{35}K^2\;\;,\;\;
\widetilde{\delta}_2(\tau_w)  = \widetilde{\delta}_3(\tau_w) =
\widetilde{\delta}_4(\tau_w)  = 0\;\;,
\end{equation}
where $K\ge0$.  Inserting these functions into the radial expansion
coefficients in the solution for the $\ms{C}_n,\;n\in\{1,\,2,\,3\}$
corrections in section \ref{sec:vac_sol}, and comparing the thus
obtained expressions with the radial expansion of the static,
axially-symmetric vacuum solution,  so-called SIMPLE 
(see Bi{\v c}{\'a}k \etal 1983, G{\'o}mez, \etal 1994 or \ref{app:SIMPLE}),
in \eqref{eq:exp_gamma_simp} - \eqref{eq:exp_Phi_simp} shows that they
agree.  Hence, putting the functions \eqref{eq:free_SIMPLE} into our
general solution for $\ms{C}_1,\,\ms{C}_2$ and $\ms{C}_3$ approximates
SIMPLE in a sufficiently small neighborhood of the vertex. Moreover,
using the Bondi--Sachs metric at the vertex determined by our first
approach, we can identify $K$ with the Riemann normal component
$6\times \mc{R}_{1313}$, which gives rise to  the $(l=2) -$ multipole of the electric part of the Weyl tensor.

\begin{table}
\caption{\label{tab:timedriv_hierachy} The hierarchical dependence of the time derivatives of the
  free functions arising in the solution of the correction coefficients $\ms{C}_1,\, \mc{C}_2$, and $\ms{C_3}$, where the time derivatives with respect to $\tau_w$ are denoted by overdots.}
\begin{indented}
\item[]
\begin{tabular}{ccc}
\hline
correction & free  & time  \\[-3ex]
coefficient & functions & derivatives \\
\hline
$\ms{C}_1$ & $\widetilde{\gamma}_2(\tau_w),\;\widetilde{\delta}_2(\tau_w)$ &$\emptyset$ \\
$\ms{C}_2$ & $\widetilde{\gamma}_3(\tau_w),\;\widetilde{\delta}_3(\tau_w)$ & $\dot{\widetilde{\gamma}}_2(\tau_w)$, $\dot{\widetilde{\delta}}_2(\tau_w)$\\
$\ms{C}_3$ & $\widetilde{\gamma}_4(\tau_w),\;\widetilde{\delta}_4(\tau_w)$ & $\dot{\widetilde{\gamma}}_{3}(\tau_w)$, $\ddot{\widetilde{\gamma}}_{2}(\tau_w)$, 
$\dot{\widetilde{\delta}}_{3}(\tau_w)$, $\ddot{\widetilde{\delta}}_2(\tau_w)$\\
\hline
\end{tabular} 
\end{indented}
\end{table}
By linearizing the vacuum Einstein equations with respect to the
coefficients $\ms{C}_n, \,1\le n\le N,$ of $\ms{F}$, we found how the
hierarchical pattern of the time derivatives can be determined for
arbitrary $N$, which provided us with the complete boundary conditions at a regular vertex for
the Bondi--Sachs metric in linearized gravity and for axially symmetric
vacuum space-times. The solution depends on $2N$
free functions, $\widetilde{\gamma}_n$ and $\widetilde{\delta}_n$,
along the geodesic $\bs{c}(\tau)$.  Regularity of the vertices along
$\bs{c}(\tau)$ is guaranteed, if the two free functions in the
coefficient $\ms{C}_n$ are at least $(N-n)$ times differentiable with
finite derivatives along $\bs{c}(\tau)$. These $2N$ functions also
determine the complete axially, symmetric initial data on the null
cone for the linearized Bondi--Sachs metric {\it in vacuo}. We claim
that these $2N$ functions also determine the complete non-linear
initial data on the cone to arbitrary order of $N$. Up to $N=3$, this
claim is backed up by our calculations. For $N>3$, we learn from the
coupling of the Riemann normal components in the Taylor series of
normal coordinates at a space-time point (Schouten, 1954, Thomas,
1991) that the non-linear coupling in a higher order coefficient
occurs only between Riemann normal components and covariant
derivatives determining lower order coefficients. Hence, our claim
yields the answer to question (IV) raised in the introduction. We
formulate it as a \textbf{Conjecture: }
\begin{quote}
 {\it Suppose $(\ms{M}^4,\,g)$ is an axially symmetric vacuum
   space-time, $\bs{c}(\tau)$ a time-like geodesic on the axis of
   symmetry $\ms{A}$, $\ms{N}_w$ is a future ${(w=1)}$ or past
   ${(w=-1)}$ null cone with its vertex $\ms{O}$ on $\bs{c}(\tau)$,
   $g(x^\alpha)$ is a Bondi--Sachs metric on $\ms{N}_w$ with respect
   to coordinates $x^\alpha=(\tau_w,\,r,\,y,\,\phi)$, and its six
   metric functions $\ms{F}$ are represented by the finite power
   series \eqref{eq:ser_gamma}-\eqref{eq:ser_UA} in terms of the areal
   distance $r$ with coefficients $\ms{C}_n,\,1<n\le N$ evaluated at
   $\ms{O}$. Then the initial data on $\ms{N}_w$ for $\ms{F}$ are
   given by $2N$ functions along $\bs{c}(\tau)$, which can be chosen
   freely. Regularity at $\ms{O}$ requires that $\ms{C}_n$ must be
   decomposed by a spectral series of associated Legendre polynomial
   $P^{m}_{n+1}(y)$ with respect to the angular variable $y$ and that
   the two free functions in the coefficient $\ms{C}_n$ are at least
   $(N-n)$ times differentiable and the derivatives of these functions
   are finite along $\bs{c}(\tau)$.  }
\end{quote}

\noindent
The main consequence of the conjecture concerns the arbitrariness of
the vacuum initial data that can be imposed on a null cone in
Bondi-Sachs coordinates. If the null cone is assumed to have a vertex
and if this vertex is assumed to be a regular point in the null cone,
then one cannot impose any data. Instead the data have to obey a
regular angular structure as it is provided by the solution of the
Einstein equations for the coefficients of a Taylor series expansion
of the metric at the vertex. The only freedom one has left in the
choice of the data are free functions on the geodesic tracing the
vertex. These free functions determine the highest $l-$multipole of
the Legendre basis $P^m_l$ in each power of $r$ in the Taylor series
of the initial data at the vertex. They correspond to the $l-$multipoles of the axisymmetric parts of the electric and magnetic parts of the Weyl tensor evaluated along the time-like geodesic.

Our study raises further questions which should be investigated such
as: How many free functions determine the initial data on a null cone
with a regular vertex in space-times with no symmetry?  What are the
properties these free functions have to obey so that the initial data
on the null cone describe an asymptotically flat space-time? What is
the structure of the Bondi--Sachs metric at the vertex, if matter is
assumed at the vertex and in its neighborhood?  We plan to address the
last question in an future publication.
\ack
\begin{appendix}
We are very grateful for discussions with B. Schmidt, H. Friedrich,
 A. Bauswein, E. Gourgoulhon, and Y. Choquet-Bruhat. We thank
 I. Cordero-Carrion, P. Jofr{\'e} Pfeil, J. Winicour and L. Lehner for
valuable comments on the manuscript and A. J. Penner for providing
grammatical corrections. Financial Support is acknowledged from the
Collaborative Research Center on Gravitational Wave Astronomy of the
Deutsche Forschungsgemeinschaft (DFG SFB/Transregio 7), from the Max
Planck Society, the Bundesministerium f{\"u}r Familie, the observatory of Paris, the university of Paris Diderot and the CNRS.
\section{Tensor components for the Einstein equations}
\label{app:Einstein}
In the following, we give the components of
$\mc{R}_{\alpha\beta}$ that are used in Sect.\,\ref{sec:sols} to
calculate the solution of the vacuum Einstein equations up to
third-order corrections and for the solution of the $n^{th}-$order
correction coefficients of linearized vacuum Einstein equations. We
also introduce the following notations
\begin{eqnarray}
\fl  \mc{R}_{(2D)}    &:=& g^{AB}\mc{R}_{AB}\;\;, \\
\fl  \mc{R}_{(\gamma)} &:=& \Big[s^2\rme^{ -2\gamma}\mc{R}_{yy} 
   + s^{-2}\rme^{ 2\gamma}\mc{R}_{\phi\phi}\Big]\cosh^{-1}(2\delta)\;\;,\\
\fl  \mc{R}_{(\delta)} &:=& \Big[s^2\rme^{ -2\gamma}\mc{R}_{yy} 
   - s^{-2}\rme^{ 2\gamma}\mc{R}_{\phi\phi}\Big]\sinh(2\delta) 
   - 2\cosh(2\delta)\mc{R}_{y\phi}\;\;.
\end{eqnarray}
\subsection{Ricci Tensor Components for the Third-Order 
            Correction Coefficients}
The Ricci tensor components for the supplementary conditions are
\begin{eqnarray}
\fl   R_{\tau_w\tau_w}& = & 
   \Big[s^2(\Phi_2+2\beta_2)_{,y}\Big]_{,y}+ 6(\Phi_2+2\beta_2)
   +\bigg\{   \Big[s^2(\Phi_3+2\beta_3)_{,y}\Big]_{,y}+ 12(\Phi_3+2\beta_3)\nonumber\\
 \fl &&-4\beta_{2,\tau_w}+2\Phi_{2,\tau_w}-U^y_{1,y}\bigg\}wr
   +\bigg\{ \Big[s^2(\Phi_4+2\beta_4)_{,y}\Big]_{,y}+ 20(\Phi_4+2\beta_4)\nonumber\\
\fl &&   +2\Phi_{3,\tau_w}-6\beta_{3,\tau_w}  -U^y_{2,y\tau_w}
             + (2\Phi_2+4\beta_2-2\gamma_2)\Big[s^2(\Phi_2+2\beta_2)_{,y}\Big]_{,y} \nonumber\\
\fl  &  & 
         +\Big[4yU^\phi_1U^\phi_{1,y} -2 (\Phi_{2,y}+2\beta_{2,y})\gamma_{2,y}
         +2(\Phi_{2,y})^2+6(\beta_{2,y})^2    +  6\Phi_{2,y}\beta_{2,y}\nonumber\\
\fl &&  -\f{9}{2}(U^\phi_1)^2\Big]s^2
          -\Big[U^\phi_1U^\phi_{,yy}+\f{1}{2}(U^\phi_{1,y})^2\Big]s^4
           -(\Phi_{2,y}+U^y_{1,yy}+8\beta_{2,y})U^y_1 \nonumber\\
\fl &&          
        +(96\beta_2-2U^y_{1,y})\Phi_2
         +32\Big[(\Phi_2)^2 +2(\beta_2)^2\Big]
         -\Big[2y U^y_{1,y}+\f{13}{2}U^y_1\Big]s^{-2}U^y_1\nonumber\\
\fl &&
        -4\beta_2 U^y_{1,y}  -(U^y_{1,y})^2 -2s^{-4}y^2(U^y_{1})^2
          \bigg\}r^2+\mc{O}(r^3)      
\end{eqnarray}
\begin{eqnarray}
\fl  r^{-1}s^2\mc{R}_{y\tau_w}&=&
       \Big[2(\beta_2+\Phi_2)_{,y}s^2 +3U^y_{1}\Big]w
       +\Big\{\Big[3(\beta_3+\Phi_3)-\beta_{2,\tau_w} \big]_{,y}s^2 +6U^y_{2}\nonumber\\
\fl &&+(s^2\gamma_{2,\tau_w})_{,y}-\f{1}{2}U^y_{1,\tau_w}\Big\}r
      +\bigg\{\Big[4(\Phi_4+\beta_4)+\beta_{3,\tau_w}\Big]_{,y}s^2
      +(s^2\gamma_{3,\tau_w})_{,y}\nonumber\\
\fl  &&
      +(s^2\beta_{2,y})_{,y}U^y_{1} +4\Big[(2\Phi_2+3\beta_2-\gamma_2)(\Phi_2+\beta_2)_{,y} +4(3U^\phi_1-yU^\phi_{1,y})\delta_2   \Big]s^2\nonumber\\
\fl   &&
      +\Big[\delta_2 U^\phi_{1,yy}-2U^\phi_1U^\phi_{1,y}\Big]s^4 
      +\Big(10 \Phi_2+2\beta_2+14\gamma_2-\f{11}{2}U^y_{1,y}\Big)U^y_1+10U^y_3\nonumber\\
\fl &&
      -U^y_{2,\tau_w}-4ys^{-2}(U^y_1)^2\Big\}wr^2+\mc{O}(r^3)        \\
\fl    r^{-1}s^{-2}\mc{R}_{\phi\tau_w}&=&    
        \Big[\f{1}{2}s^{-2}(s^4 U^\phi_{1,y})_{,y}+2U^\phi_1\Big]w
        +\Big\{\Big[\f{1}{2}(s^4 U^\phi_{2,y})_{,y}+ (s^2\delta_{2,\tau_w})_{,y} \Big]s^{-2}+5U^\phi_2
\nonumber\\
\fl &&        
        -\f{1}{2}U^\phi_{1,\tau_w}  \Big]r  
           +\bigg\{    \Big[\f{1}{2}(s^4 U^\phi_{3,y})_{,y}+(s^2\delta_{3,\tau_w})_{,y}-(s^4\gamma_2)_{,y}U^\phi_1-2(s^4U^\phi_{1,y})_{,y}\gamma_2\Big]s^{-2}
           \nonumber\\
\fl&&    +9U^\phi_3
          -U^\phi_{2,\tau_w}-2yU^\phi_1\beta_{2,y}-2s^2U^\phi_{1,y}\gamma_{2,y} 
          +\Big(2\beta_2-\f{3}{2}U^y_{1,y}-10 \gamma_2 +10 \Phi_2 \Big)U^\phi_1  \nonumber\\
\fl && +\Big(\delta_{2,yy}-2U^\phi_{1,y}\Big) U^y_1 +\Big[U^y_{1,yy}-4(\Phi_2+\beta_2)_{,y}\Big]\delta_2        
          +2\Big[2yU^\phi_1 +7\delta_2\Big]s^{-2} U^y_1\nonumber\\
\fl && +2\delta_{2,y}U^y_{1,y}\bigg\}+\mc{O}(r^3)          
\end{eqnarray}
The Ricci tensor components for the hyper-surface equations are
\begin{eqnarray}
\fl  \mc{R}_{rr}  &=&  8\beta_2  + 12\beta_3 w r+\Big[16\beta_4-8(\gamma_2^2+\delta_2^2)\Big]r^2+\mc{O}(r^3)\;\;,\label{eqapp:R_rr} \\
\fl   r^{-1}s^2\mc{R}_{ry} & =&
                    2U^y_1+2(s^2\gamma_2)_{,y} 
                     +\Big[5U^y_2-s^2\beta_{3,y}+3(s^2\gamma_3)_{,y}  \Big]r w^{-1}                    
                  \nonumber\\
\fl                    &&
 +\Big[9U^y_3-2s^2\beta_{4,y}
                   -4s^2\gamma_2\gamma_{2,y}
                  + 4(s^2\gamma_2)_{,y} +(6U^\phi_1 -4\delta_{2,y})s^2\delta_2 
                  \nonumber\\
\fl              &&
              +6(\gamma_2-\beta_2)U^y_1\Big]r^2
              +\mc{O}(r^3)\;\;,\label{eqapp:R_ry}
\end{eqnarray}              
\begin{eqnarray}
\fl   r^{-1}s^{-2}\mc{R}_{r\phi}&=&              
                    2U^\phi_1+2s^{-2}(s^2\delta_2)_{,y} 
                    +\Big[5U^\phi_1 +3s^{-2}(s^2\delta_2)_{,y}  \Big]r w^{-1}                    
                  \nonumber\\
\fl                 &&
                 +\Big[9U^y_3
                 + 4s^{-2}(s^2\delta_4)_{,y} 
                 +6s^{-2}U^y_1 \delta_2 
                 -6(\gamma_2+\beta_2)U^\phi_1\Big]r^2
              +\mc{O}(r^3)\;\;,\label{eqapp:R_rphi}
\end{eqnarray}              
\begin{eqnarray}              
\fl   \mc{R}_{(2D)}   
  &=&
  -12\Phi_2  + 5U^y_{1,y} -2\big(\beta_{2,y}s^2\big)_{,y}-8\beta_2 
     +2s^{-2}\big(s^4\gamma_{2,y}\big)_{,y}-4\gamma_2    \nonumber\\
\fl     &&      
     +\Big[-16\Phi_3 +6U^y_{2,y}
               -2\big(\beta_{3,y}s^2\big)_{,y}-12\beta_3
               +2s^{-2}\big(s^4\gamma_{3,y})_{,y}-4\gamma_3 \Big]wr\nonumber\\
\fl && +\bigg\{-20\Phi_4+7U^y_{3,y}-2\big(\beta_{4,y}s^2\big)_{,y}-16\beta_4         
             +2s^{-2}\big(s^4\gamma_{4,y}\big)_{,y}-4\gamma_4 \nonumber\\
\fl && 
            \quad -4s^2(\gamma_2^2+\delta_2^2)
            -20\Phi_2^2
            -\f{1}{2}\Big(\f{U^y_1}{s}\Big)^2 
            -\f{1}{2}(sU^\phi_1)^2
                   -4s^{-2}(s^4\delta_{2,y})\delta_2
    \nonumber\\
\fl && \quad
       -2s^2\big(\beta_{2,y}-2\gamma_{2,y}\big)\beta_{2,y}
            +4\Big[(s^2\beta_{2,y})_{,y}-s^{-2}(s^4\gamma_{2,y})_{,y}\Big]\gamma_2
\nonumber\\
\fl && \quad       
       +4(\gamma^2_2+\delta_2^2)
       -2(8\Phi_2+5U^y_{1,y})\beta_2
             \Big]\bigg\}r^2    +\mc{O}(r^3)  \;\;, \label{eqapp:R_2d}
\end{eqnarray}             
and for the evolution equations
\begin{eqnarray}             
\fl    r^{-2}\mc{R}_{(\gamma)}  
  &=&
        -12\gamma_2 -2s^2\beta_{2,yy}  +3s^{2}\big(s^{-2}U^y_1)_{,y}
        +\Big[12\gamma_{2,\tau_w}-24\gamma_3 -2s^2\beta_{3,yy} \nonumber\\
 \fl&& +4s^{2}\big(s^{-2}U^y_2)_{,y}\Big]wr
     +\bigg\{
      16\gamma_{3,\tau_w} -40\gamma_4+5 s^{2}\big(s^{-2}U^y_3)_{,y} -2s^2\beta_{4,yy}\nonumber\\
      \fl && -6s^2\beta_2\big(s^{-2}U^y_1)_{,y}
       +4\Big(U^y_{1,y}-10\Phi_2-4\beta_2\Big)\gamma_2        +14U^y_1\gamma_{2,y}-\f{1}{2}\Big(\f{U^y_{1}}{s}\Big)^2       \nonumber\\
\fl&&
       +\Big[4\gamma_2\beta_{2,yy}-2\beta_{2,y}^2+14\delta_2U^\phi_{1,y}  +\f{1}{2}(U^\phi_1)^2\Big]s^2\bigg\}r^2
       +\mc{O}(r^3)\;\;,\label{eqapp:R_gamma}\\
\fl   r^{-2}\mc{R}_{(\delta)}  &=&
           12\delta_2-3s^2U^\phi_{1,y}    
           +\Big(-12\delta_{2,\tau_w}+24\delta_3-4s^2U^\phi_{2,y}\Big)wr
          \nonumber\\
\fl && 
 +\bigg\{
            -16\delta_{3,\tau_w}+40\delta_4          
        -5s^2U^\phi_{3,y}
        +\Big[(6\beta_2 +14\gamma_2)U^\phi_{1,y}-4\beta_{2,yy}\delta_2\Big]s^2\nonumber\\
\fl && \quad       
        +(U^\phi_1-14\delta_{2,y})U^y_1
       +4(4\beta_2-U^y_{1,y}+10\Phi_2)\Big]\delta_2\bigg\}r^2+\mc{O}(r^3)\;\;.\label{eqapp:R_delta}
\end{eqnarray}
\subsection{The linearized Ricci tensor for the $\ms{C}_n$ 
            correction coefficients}
\label{app:lin}
The Ricci quantities for the main equations that are linearized with
respect to $\ms{C}_n$ are assumed to be of the form
\begin{displaymath}
\fl  \mc{R}_{\mu\nu} = \sum_{n=0}^N (wr)^n \mc{R}^{(n)}_{\mu\nu}\;\;,\;\;
 \mc{R}_{(\cdot)} = \sum_{n=0}^N \Big(\f{r}{w}\Big)^n \mc{R}^{(n)}_{(\cdot)}\;\;,\;\;
\end{displaymath}
where $\mc{R}_{(\cdot)}\in \{\mc{R}_{(\gamma)},\, \mc{R}_{(\delta)},\,
\mc{R}_{(2D)}\}$. The relevant non-zero coefficients for the
hyper-surface equations read
\begin{eqnarray}
\fl  \mc{R}^{(n)}_{rr}&= (4n+8)\beta_{n+2} &: n\ge0\;\;,\label{eq:Rrr_lin} \\
\fl   s^{2}\mc{R}^{(n)}_{ry}&  =   \f{1}{2}n(n+3)U^y_n+ (1-n)\beta_{n+1,y} +(n+1)\Big(s^2\gamma_{n+1}\Big)_{,y}\qquad &:n\ge 1\;\;,\label{eq:Rr2_lin}\\
\fl   s^{-2}\mc{R}^{(n)}_{r\phi} & =   \f{1}{2}n(n+3)U^\phi_n +s^{-2}(n+1)\Big(s^2\delta_{n+1}\Big)_{,y} &:n\ge 1\;\;,\label{eq:Rr3_lin}\\
\fl   \mc{R}^{(n)}_{(2D)}& = -4(n+3)\Phi_{n+2}
                                          -4(n+2)\beta_{n+2}
                                           +(n+5)U^y_{n+1,y}
                                            \nonumber\\
\fl    &\quad-2\Big(s^2\beta_{n+2,y}\Big)_{,y}                                 
             +2s^{-2}\Big(s^4\gamma_{n+2,y}\Big)_{,y}
                                           -4\gamma_{n+2} &:n\ge 0\;\;,        \label{eq:R2d_lin}  
\end{eqnarray}
and the evolution equations for $\gamma$ and $\delta$ are given by
\begin{eqnarray}
   \fl \mc{R}^{(2)}_{(\gamma)} &= -12\gamma_2 -2s^2\beta_{2,yy}
                                                        + 3 s^2
\Big(s^{-2}U^y_1\Big)_{,y}\;\;,\label{eq:Rg_lin_2}
\\
    \fl \mc{R}^{(2)}_{(\delta)} &=  12\delta_2 - 3 s^2 U^\phi_{1,y}\;\;,
\label{eq:Rd_lin_2}
\end {eqnarray}
and by
\begin{eqnarray}
    \fl \mc{R}^{(n+1)}_{(\gamma)} &=
         2(n+1)\Big[2\gamma_{n,\tau_w} - (n+2)\gamma_{n+1}\Big]
         + s^2(n+2) 
\Big(s^{-2}U^y_{n}\Big)_{,y}
         -2s^2\beta_{n+1,yy} 
        \;\;,\label{eq:Rg_lin_n}
\\
  \fl \mc{R}^{(n+1)}_{(\delta)} &=
         2(n+1)\Big[(n+2) \delta_{n+1} - 2\delta_{n, \tau_w}\Big]
          -s^2(n+2) U^\phi_{n,y}\;\;.
\label{eq:Rd_lin_n}
\end {eqnarray}
for $n>1$,  respectively.
\section{The vacuum solution SIMPLE}\label{app:SIMPLE}
The static vacuum solution SIMPLE is the only known explicit
non-linear and axially symmetric solution of the Einstein equations in
Bondi coordinates (Bi{\v c}{\'a}k \etal 1983).  SIMPLE is boost and
rotation symmetric (Bi{\v c}{\'a}k \etal 1984), and depends on one
free parameter $a$. The solution given in G{\'o}mez \etal (1994) reads
in our coordinates $x^\alpha = (\tau_w,\,r,\,y,\,\phi)$ and
nomenclature of variables
\begin{eqnarray}
\fl   \Sigma(a,\,r,\,y)&=\sqrt{1+a^2r^2s^2(y)}\\
\fl   \gamma(r, \,y) &= \ln(1+\Sigma (a, r, y)) -\ln(2)\;\;,\;\;\\
\fl    \beta(r,\,y)&=\ln\Big[1+\Sigma(a,\,r,\,y)\Big]-\f{1}{2}\ln\Sigma(a,\,r,\,y)-\ln 2\;\;,\;\;\\
\fl   U^y(r,\,y)&= \f{a^2rys^2(y)}{\Sigma(a,\,r,\,y)}\;\;,\\
\fl   \Phi(r,\,y)&=\f{1}{2}\ln\Big[2 a^2 r^2 s^2(y) -a^2r^2 +1\Big]-\ln\Big[1+\Sigma(a,\,r,\,y)\Big]+\ln2\;\;,\\
\fl   \delta(r,\,y)&=U^\phi(r,\,y)=0\;\;.
\end{eqnarray}
Defining $a= (12K )^{1/2} $ and expanding the Bondi functions of
SIMPLE near $r=0$ gives
\begin{eqnarray}
\fl  \gamma(r,\,y) & =  r^2KP^2_2(y) + \bigg[-\f{27}{2}K^2P^2_2(y)+\f{9}{35}K^2P^2_4(y)\bigg]r^4+\mc{O}(r^6)\;\;,\label{eq:exp_gamma_simp} \\
\fl   \delta(r,y)&=0\;\;,\label{eq:exp_delta_simp}\\
\fl  \beta(r,\,y)&= \bigg[\f{9}{7}K^2P^2_2(y)-\f{3}{35}K^2P^2_4(y)\bigg]r^4+\mc{O}(r^6)\;\;, \\
\fl   U^y(r,\,y) & = -4rKs(y)P^1_2(y)+ \bigg[\f{96}{7}K^2P^1_2(y)-\f{144}{35}K^2P^1_4(y)\bigg]s(y)r^3+\mc{O}(r^5), \\
\fl  U^\phi(r,y)&=0\;\;,\\
\fl  \Phi(r,\,y)&= -6r^2KP^0_2(y) + \bigg[-\f{48}{5}K^2 +\f{24}{7}K^2P^0_2(y)-\f{1044}{35}K^2P^0_4(y)\bigg]r^4+\mc{O}(r^6)\;\;,\nonumber\\
\fl&\label{eq:exp_Phi_simp}
\end{eqnarray}
\end{appendix}
where we expressed powers of $y$ in terms of associated Legendre
polynomials.
\section*{References}
\begin{harvard}
\item Bondi H, van der Berg M J G and Metzner A W K, 
          1962,  
          Gravitational Waves in General Relativity. VII. Waves from Axi-Symmetric Isolated Systems, 
          \PRS {\it of London. Series} A {\bf 269} 21--52
\item Sachs R K, 
          1962, 
          Gravitational Waves in General Relativity. VIII. Waves in Asymptotically Flat Space-Time, 
          \PRS {\it of London. Series} A {\bf 270} 103--126
\item  Newman E T and Unti, 
          1962,
          Behavior of Asymptotically Flat Empty Spaces,
          \JMP {\bf 3} 891--901
\item Newman E T and Penrose R, 
         1962, 
         An Approach to Gravitational Radiation by a Method of Spin Coefficients,
         \JMP {\bf 3},  566--79 
\item   Newman E T and Penrose R, 
         1963, 
        Errata: An Approach to Gravitational Radiation by a Method of Spin Coefficients,
         \JMP {\bf 4}, 998
\item Tamburino L and Winicour J, 
         1966, 
         Gravitational Fields in Finite and Conformal Bondi Frames, 
         \PR  {\bf 150} 1039--1053
\item Bondi H, 1960, 
	Gravitational Waves in General Relativity,
	{\it Nature}, {\bf 189}, 535
\item Isaacson R A, Welling J S, Winicour J,
	1983,
	Null cone computation of gravitational radiation,
	\JMP, {\bf 24}, 1824--1834	
\item G{\'o}mez  R, Papadopoulos  P, Winicour  J,
         1994,
         Null cone evolution of axisymmetric vacuum space-times,
         \JMP, {\bf 35}, 4184--4204
\item Winicour J, 
          2012a,         
          Characteristic Evolution and Matching,
          {\it Living Review in General Relativity}, {\bf 15}
\item Freitag E, Busam R, 2005,
          Complex Analysis, Springer, p. 547	          
\item Winicour J, 
          2012b,         
          private communication
\item Thorne K, 1980, Multipole expansions of gravitational radiation, {\it Rev. Mod. Phys}, {\bf 52}, 299
\item Thorne K, Hartle J, 1985, Laws of motion and precession for black holes and other bodies, \PR D, {\bf 31}, 1815
\item Zhang X, 1986,  Multipole expansions of the general-relativistic gravitational field of the external universe,  \PR D, {\bf 34}, 991                              
\item Siebel F,  Font J A,  M\"uller E, Papadopoulos P,
         2002,
         Simulating the dynamics of relativistic stars via a light-cone approach,
         \PR D, {\bf 65}, 064038         
\item Penrose R,
          1963
         Null hypersurface initial data for classical fields of arbitrary spin and for general relativity,
         {\it  Aerospace Research Laboratories (P.G. Bergmann)}, p.   63--56,
         republished in {\it Gen. Rel. Grav}, {\bf 12 }, p. 225--256
\item Friedrich H,
         1986,
         On purely radiative space-times,
         {\it Comm. Math. Phys.} {\bf103}, 35--65
\item 	
	Ellis  G F R, Nel S D, Maartens R, Stoeger W R, Whitman A P,         
	1985,
	Ideal observational cosmology,
	{\it Physics Reports}, {\bf 124}, 315--417
\item Poisson E, 
         2004,
         Retarded coordinates based at a world line and the motion of a small black hole in an external universe,
         \PR D , {\bf 69}, 084007
\item Poisson E, 
         2005,
         Metric of a Tidally Distorted Nonrotating Black Hole,
         \PRL , {\bf 94}, 0161103
\item Preston B, Poisson E, 
         2006a,
         Light-cone coordinates based at a geodesic world line,
         \PR D , {\bf 74}, 064009
\item Preston B, Poisson E, 
         2006b,
          Light-cone gauge for black-hole perturbation theory,
         \PR D , {\bf 74}, 064010  
\item Poisson E,  Vlasov I,
         2010,
         Geometry and dynamics of a tidally deformed black hole,
         \PR D, {\bf 81}, 024029
\item Choquet--Bruhat Y, Chu{\'s}ciel P T, Mart{\'i}n-Carc{\'i}a J M,
         2010,
         An existence theorem for the Cauchy problem on the light-cone for the vacuum Einstein equations with near-round analytic data,
         {\it Preprint : qr-qc/1012.0777}              
\item Choquet--Bruhat Y, Chru{\'s}ciel P T, Mart{\'i}n-Garc{\'i}a J M	,
         2011, 
         The Cauchy Problem on a Characteristic Cone for the Einstein Equations in Arbitrary Dimensions,
         {\it Annales Henri Poincar{\'e}}, {\bf 12},  419--482
\item  Chru{\'s}ciel P T, Jezerski, J, 
          2010, 
          On free general relativistic initial data on the light cone, 
          {\it Preprint gr-qc/1010.2098v1}
\item Chru{\'s}ciel P T, Paetz, T, 
         2012,
         The many ways of the characteristic Cauchy problem,
          {\it Preprint gr-qc/1203.4534v1}
\item P. Chru\'sciel, J. Jezierski and M. Maccallum,  \PR D,  {\bf 58}, 84001,  (1998)
\item P. Chru\'sciel, J. Jezierski and J. Kijowski, Lecture Notes in Physics  {\bf 70}, Springer-Verlag (2002)
\item Bardeen J, Piran T,
          1983,
          General relativistic axisymmetric rotating systems: coordinates and equations,
          {\it Physics Reports}, {\bf 96}, 205 - 250
\item Misner Ch, Thorne K S , Wheeler J A, 
         1973,
         Gravitation,           
         San Francisco, W. H. Freeman, p. 1215
\item Jackson J D,
        1999,
        Classical Electrodynamics,
        John Wiley  and Sons, Inc.,  p. 808 
\item Dautcourt G,
         1967,
         Characteristic Hypersurfaces in General Relativity I,
         \JMP, {\bf{8}}, 1492--1501          
\item Thomas T Y,
         1991,
         The Differential Invariants of Generalized Spaces,
         American Mathematical Society, $2^{nd}$ edition, p. 240
\item Schouten J A,
         1954,
          Ricci-Calculus: An Introduction to Tensor Analysis and its Geometrical Applications,
          Springer, $2^{nd}$ edition, p. 540                       
\item Misner C W, Manasse F K,
        1963,
        Fermi Normal Coordinates and Some Basic Concepts in Differential Geometry,
        \JMP, {\bf 4}, 735--745   
\item Iliev B Z, 
         2006,
         Handbook of Normal Frames and Coordinates,
         Birkh{\"a}user, Series: Progress in Mathematical Physics, p. 441
\item Newman E, Posadas R,
          1969,
          Motion and Structure of Singularities in General Relativity,
          \PR, {\bf 11}, 1787--1791        
\item Jordan P, Sachs R, 
          1961, 
          Beitr{\"a}ge zur Theorie der reinen Gravitationsstrahlung 
          -- Strenge L{\"o}sungen der Feldgleichungen der Allgemeinen Relativit\"atstheorie II, 
          {\it Akad. Wiss. Lit. Mainz. Abhandl. Math. Nat. Kl.} {\bf 1} 1--62         
\item Sachs R K, 
          1961, 
          Gravitational Waves in General Relativity. VI. The outgoing radiation condition, 
          \PRS {\it of London Ser.} A {\bf 264} 309--338         
\item Ni W T, Li W Q, 
         1979,
         Expansions of the affinity, metric and geodesic equations in Fermi normal coordinates about a geodesic,
         \JMP, {\bf 20}, 1925--1929   
\item Dolgov A D, Khriplovich I B,         
	1983,
	Normal coordinates along a geodesic,
	{\it  Gen. Rel. Grav.}, {\bf 15}, 1033--1041
\item Synge J L,	 
        1960,
        Relativity: The General Theory,
        North-Holland Publishing Company, Amsterdam, p. 520
\item Ishii M, Shibata M, Mino Y,
	2005,
	Black hole tidal problem in the Fermi normal coordinates,
	\PR D, {\bf 71}, 044017
\item Poisson E, Pound A, Vega I,
        2011,
        The Motion of Point Particles in Curved Spacetime,
        {\it Living Review in General Relativity}, {\bf 14}	
\item Bi{\v c}{\'a}k J, Reilly, P, Winicour J.
         1988,
         Boost-Rotation Symmetric Gravitational Null Cone Data,
         {\it General Relativity and Gravitation}, {\bf 20}, 171-181
\item Carter B,
         1070,
         The commutation property of a stationary, axisymmetric system                            
         {\it Comm. Math. Phys.}, {\bf 17}, 233--238
\item Carot J,
         2000,
         Some developments on axial symmetry,
         \CQG, {\bf 17}, 2675--2690
\item Stephani H, Kramer D, MacCallum M, Hoenselaers C, Herlt E,                  
        2003,
        Exact Solutions of Einstein's Field Equations,
        {\it Cambridge University Press,  2$^{nd}$ edition}, p. 732
\item Spivak M,
         1999,
         A Comprehensive Introduction to Differential Geometry, Vol. 3, 
         {Publish or Perish, 3$^{rd}$ Edition}, p. 314    
\item Bi{\v c}{\'a}k J,  Schmidt B G,
          1984,
          Isometies compatible with gravitational radiation,
          \JMP, {\bf 25}, 600-606       
\end{harvard}
\end{document}